%% file: main.tex
\documentclass[10pt,conference]{IEEEtran}
\IEEEoverridecommandlockouts
\input{packages}
\input{macros}

\begin{document}

% deep learning-based
% IDE
% chat

\title{Closing the Gap: A User Study on the Real-world Usefulness of AI-powered Vulnerability Detection \& Repair in the IDE}
% \title{\ben{Understanding the Impact of AI-based Vulnerability Detection \& Repair in the IDE: An Empirical Study of Developers in Practice}}
% \title{\ben{Understanding the Usefulness of AI-based Vulnerability Detection \& Repair in the IDE: An Empirical Study in Software Development Practice}}
% \title{\ben{Understanding the Usefulness of AI-based Vulnerability Detection \& Repair in the IDE: An Empirical User Study}}
% \title{\ben{Bringing AI to the IDE for Vulnerability Detection \& Repair: An Empirical User Study}}
% \title{\new{Understanding the Usefulness of Deep Learning-based Vulnerability Detection \& Fixing Tools \wei{in the Real-World Deployment Setting}}}
% \title{\new{Understanding the Real-World Impact of Deep Learning-based Vulnerability Detection \& Fixing Tools in Software Engineering Practice}}
% \title{Are Deep Learning-based Vulnerability Detection \& Fixing Tools Useful in Practice?}
% \title{Is Deep Learning Useful in Real-World Vulnerability Detection and Fixing?}
%\title{Closing the Gap: A User Study of Deep Learning-based Vulnerability Detection + Fixing in the IDE}

% \author{Anonymous for Review}
\author{
\IEEEauthorblockN{Benjamin Steenhoek\textsuperscript{1}\thanks{\textsuperscript{1} Work primarily done during an internship at Microsoft.}\thanks{\textsuperscript{\ \ } Corresponding email: \href{mailto:bensteenhoek@microsoft.com}{bensteenhoek@microsoft.com}}\IEEEauthorrefmark{1},
Kalpathy Sivaraman\IEEEauthorrefmark{2},
Renata Saldivar Gonzalez\IEEEauthorrefmark{2},
Yevhen Mohylevskyy\IEEEauthorrefmark{2},\\
Roshanak Zilouchian Moghaddam\IEEEauthorrefmark{2}, and
Wei Le\IEEEauthorrefmark{1}}
\IEEEauthorblockA{\IEEEauthorrefmark{1}Department of Computer Science, Iowa State University, Ames, IA, USA}
\IEEEauthorblockA{\IEEEauthorrefmark{2}Microsoft Data \& AI, Redmond, WA, USA}}

\maketitle

\input{abstract}


\input{content}

\section{Acknowledgments}

This research was partially supported by the U.S. National Science Foundation (NSF) under
Awards \#1816352 and \#2313054.
We thank Drs. Sarah Fakhoury, Christopher Bird, Denae Ford, and Thomas Zimmerman for their insightful discussions and valuable contributions to the validation of the user study design. We are also deeply appreciative of the participants of the user study for their time and input, which were essential to this research.

% \newpage
\bibliographystyle{plainnat}
% \interlinepenalty=10000
% \bibliographystyle{plainurl}
% \bibliographystyle{hunsrtnat}
\bibliography{main}

\end{document}

%% file: packages.tex
\usepackage{cite}
\usepackage{amsmath,amssymb,amsfonts}
\usepackage{algorithmic}
\usepackage{graphicx}
\usepackage{textcomp}
\usepackage[dvipsnames]{xcolor}
\usepackage[square,numbers]{natbib}
\usepackage{url}
\usepackage{booktabs}
\usepackage{multirow}
\usepackage{makecell}
\usepackage{lipsum}
\usepackage[frozencache,cachedir=.]{minted}
\usepackage{xcolor}
\usepackage{subcaption}
\usepackage{dblfloatfix}
\usepackage{tcolorbox}
\usepackage[normalem]{ulem}
\usepackage[hidelinks,pdftex,
            pdfauthor={Benjamin Steenhoek, Kalpathy Sivaraman, Renata Saldivar Gonzalez, Yevhen Mohylevskyy, Roshanak Zilouchian Moghaddam, and Wei Le},
            pdftitle={Closing the Gap: A User Study on the Real-world Usefulness of AI-powered Vulnerability Detection \& Repair in the IDE},
            pdfkeywords={deep learning, vulnerability detection, vulnerability repair, IDE, user study}]{hyperref}
\usepackage[nameinlink]{cleveref}

%% file: macros.tex
\def\BibTeX{{\rm B\kern-.05em{\sc i\kern-.025em b}\kern-.08em
    T\kern-.1667em\lower.7ex\hbox{E}\kern-.125emX}}

\newcommand{\tool}{\textsc{DeepVulGuard}}

\newcommand{\quot}[1]{\textit{``#1''}}

\definecolor{LightGray}{gray}{0.9}

\definecolor{darkred}{HTML}{C00000}
\definecolor{lightred}{HTML}{FF8B8B}
\newtcbox{\redbox}[1][]{
    on line,
    colframe=darkred,      % Border color
    colback=lightred,    % Background color
    boxrule=0.5mm,     % Thickness of the border
    arc=0mm,           % No border radius (sharp corners)
    left=0.75mm,             % Left padding
    right=0.75mm,            % Right padding
    top=0.75mm,              % Top padding
    bottom=0.75mm,           % Bottom padding
    boxsep=0mm,           % Separation between content and border
    #1
}

\newcommand{\smallheader}[1]{\noindent \underline{#1}:}

\newcommand{\takeaway}[1]{\textbf{#1}}

%% file: abstract.tex
\begin{abstract}
% \ben{Placeholder -- we expect the abstract to take at least a quarter of a page.}
%Developers primarily use security tools during pull request (PR) reviews or periodic scans, which can result in long and costly feedback loops compared to receiving feedback during code editing. However, integrating static analysis tools at edit-time poses challenges, such as long runtimes for whole-program analysis and the requirement for compilable code snippets.
Security vulnerabilities impose significant costs on users and organizations. Detecting and addressing these vulnerabilities early is crucial to avoid exploits and reduce development costs. Recent studies have shown that deep learning models can effectively detect security vulnerabilities.
Yet, little research explores how to adapt these models from benchmark tests to practical applications, and whether they can be useful in practice.
% \textcolor{gray}{\lipsum[1-3]}

This paper presents the first empirical study of a vulnerability detection and fix tool with professional software developers on real projects that they own.
% Generative Unified Analysis and Remediation of Defects
We implemented \tool{}, an IDE-integrated tool based on state-of-the-art detection and fix models, and show that it has promising performance on benchmarks of historic vulnerability data.
\tool{} scans code for vulnerabilities (including identifying the vulnerability type and vulnerable region of code), suggests fixes, provides natural-language explanations for alerts and fixes, leveraging chat interfaces.
We recruited 17 professional software developers at Microsoft, observed their usage of the tool on their code, and conducted interviews to assess the tool's usefulness, speed, trust, relevance, and workflow integration. We also gathered detailed qualitative feedback on users' perceptions and their desired features.
Study participants scanned a total of 24 projects, 6.9k files, and over 1.7 million lines of source code, and generated 170 alerts and 50 fix suggestions.
%We conducted live-usage interviews with 17 professional software developers to assess the tool's usefulness, speed, trust, relevance, and workflow integration. 
% Our study offers valuable insights and proposes actionable steps for further AI based tool development and research.
% \ben{TODO: Add some of the concrete conclusions/recommendations we make.}
We find that although state-of-the-art AI-powered detection and fix tools show promise, they are not yet practical for real-world use due to a high rate of false positives and non-applicable fixes.
User feedback reveals several actionable pain points, ranging from incomplete context to lack of customization for the user's codebase. Additionally, we explore how AI features, including confidence scores, explanations, and chat interaction, can apply to vulnerability detection and fixing. Based on these insights, we offer practical recommendations for evaluating and deploying AI detection and fix models.
% Although our model performed relatively well in on bug benchmarks, in the deployment scenario, over 50\% of false positives were caused by missing context, highlighting a need for.
Our code and data are available at this link: \url{https://doi.org/10.6084/m9.figshare.26367139}.
% \new{We release our tool's source code in our data package to support further user studies for other AI based tools.}
\end{abstract}

%% file: content.tex
\begin{figure*}[t]
    \centering
    \includegraphics[width=\textwidth]{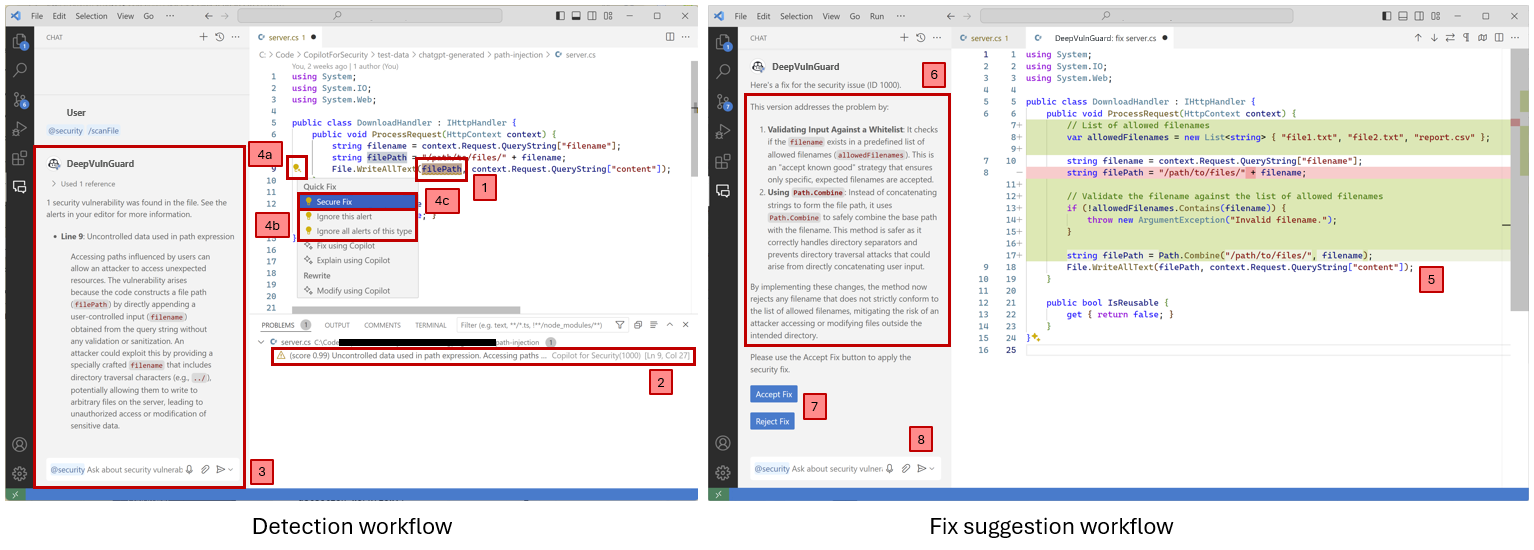}
    \caption{Overview of \tool{}'s user interface on an example program. (1) An editor alert; (2) Problems menu entry; (3) The explanation of the alert; (4a) Quick fix interaction; (4b) Ignore options; (4c) Fix trigger; (5) Suggested fix; (6) Explanation of the fix suggestion; (7) Accept/Reject buttons.
    }
    \label{fig:screenshot}
\end{figure*}

\section{Introduction}
Security vulnerabilities impact users' safety, security, and privacy and cost organizations millions of dollars per year~\cite{hbr,ibm_cost}, with reports of breaches exposing millions of records becoming commonplace~\cite{wikipediaDataBreaches}.
Early detection of vulnerabilities during the development phase can greatly reduce costs and mitigate potential impacts~\cite{book1,book2,book3}.
% To reduce security vulnerabilities, developers typically rely on static and dynamic analysis tools during pull request reviews or periodic scans. However, these methods present significant technical challenges, result in long and costly feedback loops, and can be difficult to integrate into the IDE~\cite{ohearn,edittime}.
% \ben{Need to expand a bit more by citing some statistics about impact of exploits and data breaches.}
% Traditionally, static analysis has been the primary method for identifying vulnerabilities during software development~\cite{what_to_cite_here}.
%research~\cite{those_three_highly_cited_books} indicates that the time it takes to fix a fault is positively correlated with the time between fault injection and fix.
% 
%; therefore, Static analysis is commonplace, with many major software development companies companies deploying it as part of their standard practice~\cite{staticanalysis-microsoft,staticanalysis-google,staticanalysis-facebook}.However, this approach has notable limitations, particularly when integrated into integrated development environments (IDEs). Thorough empirical studies~\cite{johnson_whydont_2013,christakis_whatdeveloperswantandneed_2016,smith_whyjohnny_2016} have highlighted several challenges associated with static analysis tools, including issues with speed, insufficient explanations, and lacking suggestions for fixes.These limitations hinder the effectiveness of static analysis in an IDE development settings, where rapid feedback and actionable insights are crucial.
% 
In recent years, deep learning (DL) vulnerability detection models have emerged as a promising approach for scanning code during software development~\cite{reveal,deepdfa,linevul}.
% \new{, and their deployment in practice could help reduce development cost}.
These models can identify vulnerability patterns in code snippets and offer the advantage of analyzing code during editing~\cite{edittime} with less configuration than traditional static analysis tools~\cite{fu_aibughunter_2023}.%\ben{is there a better citation?}.
% \ben{, as well as possibly improving on areas of speed, false positive rate}.
% \ben{Since we don't have many results specifically about the chat interface, let's re-de-emphasize it in the intro and instead highlight the role of our novel explanations.}
% 

Despite promising benchmark performance~\cite{reveal,deepdfa,linevul}, it remains unclear whether these models are actually useful in real-world development settings. %% \new{Despite promising benchmark performance~\cite{reveal,deepdfa,linevul}, 
% the practical value of these models remains unclear. They may not generalize well to new or production-quality code and have not been evaluated in the context of AI-powered user interfaces~\cite{copilot} recently integrated into the integrated development environment (IDE).}
% \ben{Do not say generalize well, we should instead say it hasn't been tested in real-world settings.}
% \wei{There have been many models developed and evaluated in the benchmark dataset. But it is not clear that whether such models can be useful in the real-world development settings.}
% \ben{Do not say generalize well, we should instead say it hasn't been tested in real-world settings.}
% While these models show promising benchmark performance~\cite{reveal,deepdfa,linevul}, they have not been evaluated in context of new AI-powered chat-based interfaces integrated into IDEs~\cite{copilot}. Also, it is not clear whether these models would be useful in practice as they may not generalize well to new or production quality code.
% \wei{however, deep learning based models may not be able to generalize to the new code, and it is not clear whether they are indeed useful in practice, and whether the developers will trust such tools. }
In the past, Major organizations such as Microsoft~\cite{staticanalysis-microsoft}, Google~\cite{staticanalysis-google}, Facebook~\cite{staticanalysis-facebook}, and Coverity~\cite{staticanalysis-coverity} have reported a gap between benchmarking success and practical application with static analyzers.
Recently, \citet{fu_aibughunter_2023} conducted a preliminary controlled study with 6 developers, showing that AI tool support reduced the time to diagnose and fix a vulnerability from 10-15 minutes to 3-4 minutes and motivating further user studies of AI detection and fix tools.
However, their study used a single bug from their dataset rather covering real-world code-bases and they only studied vulnerability detection and fixing.

In our work, we recruited 17 professional developers at Microsoft to use our detection \& repair tool in a \textit{real-world development setting} with their own projects; beyond detection and fixing, we also built and studied AI-powered \textit{explanation and chat interfaces}, which have recently become prominent in the integrated development environments (IDEs)~\cite{copilot}. Our study provides a deeper understanding of the real-world usefulness and nuances of deploying these models.

To carry out our study,
we developed \tool{}, an extension integrated with Visual Studio Code (VSCode)~\cite{vscode}, a popular IDE with over 14 million active users.
% GitHub Copilot Chat, an AI-assisted programming extension for VSCode, is also widely used, with millions of users~\cite{copilot}.
% We extended GitHub Copilot Chat to explore the effectiveness of chat-based interfaces for vulnerability detection.
% We extended GitHub Copilot Chat to explore the effectiveness of AI-powered interfaces for vulnerability detection.
% We used this extension to explore the effectiveness of chat-based interfaces for vulnerability detection.
We used state-of-the-art models, CodeBERT~\cite{codebert,edittime} and the GPT-4 large language model (LLM)~\cite{gpt4}, for detection and fix tasks.
% We recruited 17 developers from a major software company for our study;
Participants scanned 24 projects, 6.9k files, and over 1.7 million lines of source code, generating 170 alerts and 50 fix suggestions. To the best of our knowledge, ours is the first study to evaluate a detection and fix tool with professional developers on their own projects. %This reflects a real-world deployment setting for such security tools.

We initially evaluated \tool{}'s potential for deployment by testing its detection and fix models on established vulnerability datasets.
Our models achieved 80\% precision, 32\% recall, and a 46\% F1 score on SVEN~\cite{sven} for vulnerability detection and fixed 13\% of vulnerabilities on the Vul4J~\cite{vul4j2022} dataset.
\tool{} performs comparably or better than state-of-the-art models~\cite{primevul,wu_dl_apr_2023,bui_apr4vul_2023} and meets the threshold for acceptable false positives~\cite{christakis_whatdeveloperswantandneed_2016}.
These results indicate that our models are promising for detecting and fixing security vulnerabilities and can generate meaningful results for the user study.

% We interviewed the developers to understand their perceptions of the tool, including overall usefulness and aspects of speed, workflow fit, and trust, and analyzed the developers' think-aloud perceptions while using the tool, yielding insights on the individual components of detection and fix tools such as localization, confidence score, and fix suggestion.
% Our user study shows that although \tool{} could potentially be helpful (with 59\% of study participants expressing that they would use it in the future), there were several barriers to usefulness; based on the users' experiences, we provide recommendations for developing suitable vulnerability detection + fix tools.
%Our results show that \tool{} could be valuable, with 59\% of participants expressing interest in future use. 
Our results show that 59\% of participants expressed interest in future use of \tool{}, although there are several issues that limit its usefulness.
For example, one problem was an high rate of false positives in practice, caused by incorrect vulnerability pattern recognition and lack of context about code snippets (e.g., inter-procedural vulnerabilities). This highlights the need for more precise pattern recognition and better integration of environment and program context. Additionally, the requirement to trigger a manual scan significantly disrupted the users' workflow; developers prefer tools that run in the background and alert them whenever potential vulnerabilities are detected.
Regarding fixes, 75\% of proposed security fixes were unsuitable to apply ``as-is'' due to lack of customization and incorrect integration into the code. Although some fixes were functionally correct, they were not tailored to the user's codebase and could not be applied without significant modifications.
% Although sometimes a fix helped pass functional tests, it did not build on a correct approach and thus can't be used.
An interactive chat method shows promise to allow developers to guide the generation towards more applicable fixes.
% While some fixes may not be fully correct, they guide developers toward the right direction, and these partially correct fixes still offer valuable insights and encourage security best practices.
%The study also revealed that developers prefer chat interactions and background scans in detection and fix tools.
Our findings offer concrete recommendations for improving these pain points found in these tools.

In this paper, we make the following research contributions:
% \wei{I have updated the contributions below}
\begin{enumerate}
    \item We developed \tool{} a VSCode extension for detecting, explaining and fixing vulnerabilities, incorporating insights from static analysis and AI tool research. Our tool allows customization of backend models, and we provide its code in our data package to support further user studies.
    \tool{} uses a multi-task training approach for jointly predicting vulnerability classification, localization, and bug type.
    We also introduced a new vulnerability filtering method with LLMs which improved precision by over 20\%.% \roshanak{and reduced costs by Y\%}.
    % \ben{can be measured as tokens per second for snippets vs whole file.}
   
    %\item We propose a novel multi-task training approach for jointly predicting vulnerability classification, localization, and bug type.
  %  \item We introduce a new method for filtering detection classifications using large language models, which improved precision by over 20\% \roshanak{and reduced costs by Y\%}.
    \item We conducted a user study with 17 professional software engineers at Microsoft.
    Through interviews and surveys as they ran our tool on their own code, we quantitatively assessed multiple dimensions of usefulness for detection and fix tools and provided practical recommendations for improving deep learning-based vulnerability detection and fix tools.
    
   % evaluated our tool with 17 professional software engineers using their own production code bases, providing a quantitative assessment along multiple dimensions of usefulness and practical recommendations for improving deep learning-based vulnerability detection and fix tools.
\end{enumerate}

\begin{figure*}[t]
    \centering
    \includegraphics[width=\textwidth]{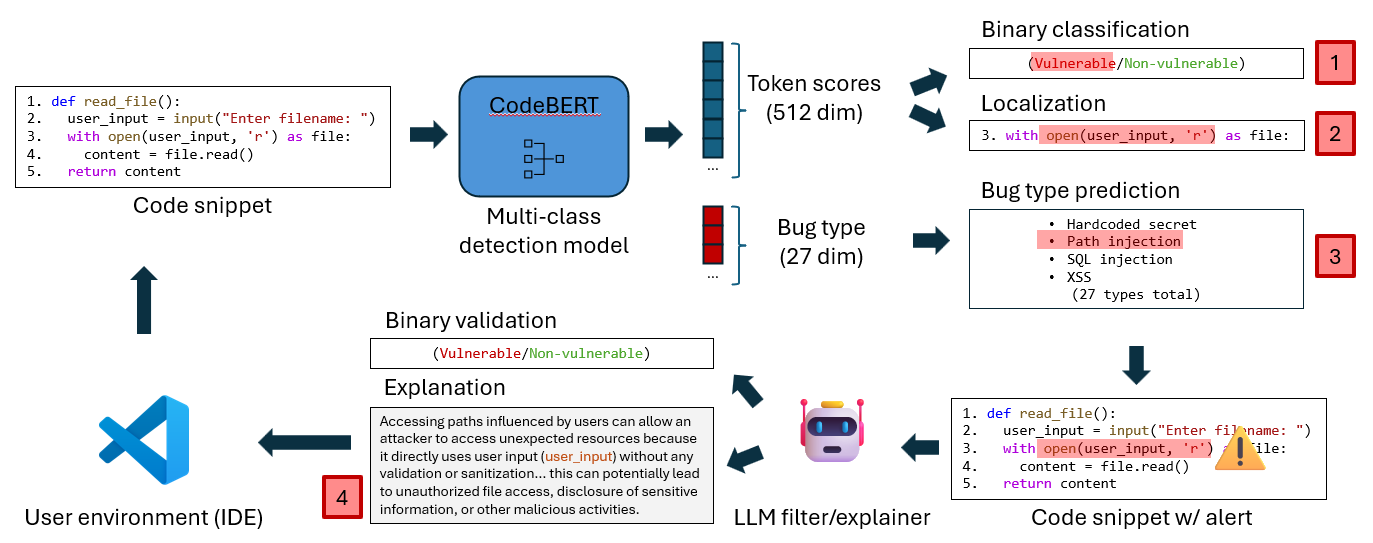}
    \caption{An overview of \tool{}'s detection workflow. (1) Binary classification into vulnerable/not-vulnerable; (2) Localization; (3) Multi-class classification into one of 27 vulnerability types; (4) Alert and explanation shown to the user.}
    \label{fig:models-overview}
\end{figure*}

% \section{Background}

% \ben{This section will be removed to focus more on our study -- all parts were distributed into the appropriate sections.}

% \ben{Deep learning-based vulnerability detection models. State-of-the-art is finetuned codebert/LLMs~\cite{linevul,diversevul,primevul}, which we use.}
% \ben{Moved to study interface section.}

% \ben{Findings and feature requests from previous studies of static analysis and AI tools~\cite{johnson_whydont_2013,christakis_whatdeveloperswantandneed_2016,fu_aibughunter_2023,wang_investigatingtrust_2023,smith_whyjohnny_2016,bird_takingflight_2023,nam_llmunderstanding_2024}. This is the motivation for our tool's user interface design and feature set.}
% \ben{Moved to study interface section.}

% \ben{How do we define ``usefulness'' of a tool and how does it compare/is it similar to prior definitions from the literature?}

% \ben{Placeholder text -- we expect intro and background to fill at least 1.5 pages.}
% \textcolor{gray}{\lipsum[1-2]}

\section{User Study Interface} %\ben{Can we be more specific?}}
% \section{Study Platform}
% \section{Study subject}

% \ben{The tool (VSCode extension). Our tool uses state-of-the-art vulnerability detection models for the IDE scenario, which we show by citing prior studies on similar architectures and through our thorough evaluation.}

% \ben{Architecture \& training of the model: codeql dataset, input/output, localization, bug type}

% \ben{Experiments: LLM-based filter, context size}

% \ben{How we evaluated the models behind the tool}

% \ben{Detection evaluation: localization metrics, break down by bug type}

% \begin{figure*}
%     \centering
%     \includegraphics[width=\textwidth, height=6cm]{example-image}
%     \caption{The user interface of \tool{} as a VSCode extension.}
%     \label{fig:models-overview}
% \end{figure*}

To study whether deep learning-based vulnerability tools can be useful in practice, we built \tool{}, a Visual Studio Code extension that brings state-of-the-art detection + fix techniques to an IDE interface. 
\tool{} allows users to (1) scan source code with CodeBERT and LLM models, (2) view the reported vulnerabilities and LLM-generated explanations directly inside the editor, and (3) generate suggestions for mitigating the vulnerability. We also implemented a telemetry module to collect user data, enabling longitudinal studies of AI-based vulnerability detection tools. As our study is the first of its kind in this area, we believe our tool will be a beneficial contribution which facilitates future user studies of vulnerability tools. We released the extension code in our data package. The code can be easily adjusted to call alternative detection and fixing solutions to be studied with developers in a real-world setting.

\subsection{IDE Integration}
\label{sec:ui}

% \ben{Show annotated screenshot of the tool and use it to explain the tool's user interface.
% % Motivate the design and feature set with other studies: ``AIBugHunter'', ``what do developers want and need'', ``trust in ai tools''.
% }

% \ben{This paragraph will include pointers to annotations on the image in the placeholders XX, pointing out the different features.}
\Cref{fig:screenshot} shows an overview of \tool{}'s user interface. Users begin by requesting to scan a file or directory. If any potential vulnerabilities arise, they are shown as highlights in the editor (1) and actionable entries in the Problems window (2), and a natural-language explanation of the vulnerability is shown in the chat panel (3).
The user can use this information to assess the vulnerability and decide if a fix is required. They can also ask questions or make suggestions to the chatbot by sending follow-up messages.
By clicking on the \textit{quick fix} lightbulb (4a), the user can ignore the specific alert or alert types (4b), or generate a \textit{quick fix} (4c).
On requesting the quick fix, the suggested code modifications will be presented in a \textit{diff} view (5), showing the lines to be removed and added. As well, an explanation of the fix is shown in the chat panel (6). The user can modify the fix in the editor or suggest improvements with natural-language chat messages if desired, then Accept it to apply it to their files or Reject it to revert to the original code (7).
Users can enter chat messages (8), e.g. asking for clarification, information, or inputs which trigger the vulnerability, and our tool will generate a conversational response.
% We enabled this feature after it was requested by several users (\Cref{sec:rq3}).

% \wei{double check and see which things we did not do. You can see this feature designed based on x, that feature design based on y.}
We drew inspiration for our tool's design from several foundational research studies on static analyzers and AI-assisted developer tools.
\citet{johnson_whydont_2013} showed that developers requested static analysis tools to be available in the IDE, along with quick fixes, and the ability to modify rule sets. Similarly, \citet{christakis_whatdeveloperswantandneed_2016} identified bad warning messages, lack of suggested fixes, and poor visualization as pain points. \citet{smith_whyjohnny_2016} presented design guidelines, such as presenting alerts in actionable locations, integrating with their workflow by tracking progress, batch processing, allowing code editing during scans, and scalability of the interface.
We incorporated all of these features into \tool{}.
% We incorporated these features, including LLM-generated explanations which provide contextual explanations of the vulnerability alerts, in our tool.
%We incorporated many of these recommended features into our tool and designed our tool to mitigate the most prevalent pain points.

A recent study on AI-powered code completion \citet{wang_investigatingtrust_2023} found that users in focus groups valued the ability to view a measure of the model's confidence. To study this in a practical implementation, we integrated confidence scores into our tool's alerts, shown in \Cref{fig:screenshot} (2).
\citet{fu_aibughunter_2023} conducted a survey study and found that most participants valued localizations, CWE type prediction, and quick fixes, so we integrated these features into our tool and evaluate them in our study, shown in \Cref{fig:screenshot} (1, 2, and 4a).
% We integrated these findings and features into our tool's feature set and investigate users' perceptions of them in our user study.
% \citet{bird_takingflight_2023}
% \citet{nam_llmunderstanding_2024}
% \cite{johnson_whydont_2013,christakis_whatdeveloperswantandneed_2016,fu_aibughunter_2023,wang_investigatingtrust_2023,smith_whyjohnny_2016,bird_takingflight_2023,nam_llmunderstanding_2024}

\subsection{Model Architecture \& Training}
% \wei{See if you can refer Figure 2 at places in your explanations}

Our tool can be easily configured to leverage a wide variety of deep learning models or static analyzers. \Cref{fig:models-overview} shows the workflow of the current design; specifically, we implemented the following techniques (please refer to our data package~\cite{our-data-package} for the implementation details, including our model training procedure, dataset statistics, and hyper-parameters). 

%We provided the desired predictions in the user interface using state-of-the-art (SOTA) models.We present a summary of each approach and document the details in an appendix inside our data package.
% \wei{key: highlight what's new/addition, the point is to show this is the state of the art, details move to the package}
% \ben{Add subtitles to show the section.}
% \ben{Summarize and move to appendix.}
% \vspace{0.1cm}
\smallheader{Fine-tuning CodeBERT for multi-task vulnerability detection} CodeBERT~\cite{codebert} and similar models consistently perform well on various vulnerability  datasets~\cite{linevul,fu_aibughunter_2023,deepdfa} with relatively low latency which is suitable for detection in the editor~\cite{edittime}.
% \ben{Why not use CodeT5, NatGen, etc as shown in DiverseVul? CodeBERT is common and simple; other models would add small performance difference with greater difficulty.}
We fine-tuned CodeBERT using \textit{multi-task learning} to (1) predict whether a code snippet contains a vulnerability, (2) localize the tokens causing it, and (3) identify the vulnerability type.
We trained on a dataset of over 1.3 million alerts labeled by CodeQL in GitHub projects following \citet{edittime}'s methodology, focusing on 27 vulnerability types related to Web security, e.g. Path Injection, SQL Injection, Hard-coded Credentials, Unvalidated URL Redirect, Cross-Site Scripting (details in data package).
We generate alerts in the extension based on the predicted vulnerability type, confidence score, and localization.

\begin{figure}[htbp]
    \centering
\begin{minted}
[
framesep=2mm,
baselinestretch=1.2,
bgcolor=LightGray,
fontsize=\footnotesize,
breaklines
]
{markdown}
You are a vulnerability detector. Only respond with "Yes" or "No" and an explanation. Does the following code snippet contain a SQL Injection vulnerability at line marked by ALERT?
\end{minted}
% \vspace{-2em}
    \caption{\tool{}'s LLM filter prompt.}
    \label{fig:filter-prompt}
\end{figure}

\smallheader{Filtering and explaining alerts with GPT-4}
% We observed that CodeBERT produced a high number of false positives (low Precision in \Cref{fig:sven-codeql} left). To mitigate these, we used GPT-4~\cite{gpt4} to filter the alerts, also generating an explanation in the same query. We annotated the code snippet with a comment describing the alert type at the localized line, e.g., for SQL injection: \texttt{// ALERT: This SQL query depends on a user-provided value}, then instructed the model: \texttt{You are a vulnerability detector. Only respond with "Yes" or "No" and an explanation. Does the following code snippet contain a SQL Injection vulnerability at line marked by ALERT?} (see appendix for full list of bug-type-specific annotations). If the answer is \textit{Yes}, the alert and LLM-generated explanation are shown to the user; otherwise, the alert is not shown.
To further filter the false positives produced by fine-tuned CodeBERT, we used GPT-4~\cite{gpt4} to filter the alerts and generate explanations. We annotated the code snippet with a comment describing the alert type at the localized line, e.g., for SQL injection: \texttt{// ALERT: This SQL query depends on a user-provided value} (see our data package for all types of annotations). Then we instructed GPT-4 to confirm whether the vulnerability is present.
If the answer is \textit{Yes}, the alert and explanation are shown to the user; otherwise, the alert is not shown ((4) in \Cref{fig:models-overview}).
We tried several prompts and evaluated on the SVEN benchmark~\cite{sven}, ultimately selecting the prompt shown in \Cref{fig:filter-prompt}. This prompt improved the Precision to an acceptable threshold of 80\%~\cite{staticanalysis-microsoft} while keeping the best Recall.

\begin{figure}[htbp]
    \centering
\begin{minted}
[
framesep=2mm,
baselinestretch=1.2,
bgcolor=LightGray,
fontsize=\footnotesize,
breaklines
]
{markdown}
A static analyzer has identified a {rule_id} security vulnerability in the {language} method below:

```
{method}
```

The SARIF result message is as follows: {message}

{description}

Write a fixed version of the method above and wrap it in triple backticks, then explain why your version addresses the problem.
\end{minted}
% \vspace{-3em}
    \caption{\tool{}'s fix model prompt.}
    \label{fig:fix-prompt}
\end{figure}

\smallheader{Prompting GPT-4 for repair and explanation}
We used GPT-4 with custom prompts to generate and explain code fixes.
The prompt, shown in \Cref{fig:fix-prompt}, includes the source code, vulnerability report, and an instruction to provide a fixed version of the code and an explanation.
% \wei{ is the prompt short enough to keep it below (see appendix)}.
We displayed the explanation in the chat panel and inserted the code suggestion to show a diff with the original content, shown in \Cref{fig:screenshot} on the right.
As with the LLM filter, we iterated on several prompts and chose the best performance on an internal dataset of bugs and Vul4J~\cite{vul4j2022}.
% \texttt{Write a fixed version of the method above and wrap it in triple backticks, then explain why your version addresses the problem} (see appendix for the full prompt).
% We parse the code suggestion and explanation from the prompt and present the explanation in the chat panel, and insert the code suggestion into the code to provide a diff with the original content.

% \wei{the following  part is not clear, please elaborate or delete it}
% \noindent{\bf Chat with GPT-4}:
% % To implement the chatbot, we passed the user's messages to GPT-4, along with a description of the tool's capabilities and the existing alerts and fix suggestions, then returned the generated response.
% We forwarded user messages to GPT-4 with a system prompt describing the tool's capabilities and previous alerts and fixes, then returned the generated response.

\subsection{Evaluating Detection and Fix Capabilities}\label{sec:benchmarking}

\begin{figure}[htbp]
    \centering
    \includegraphics[width=\linewidth]{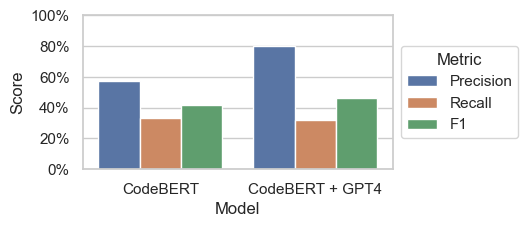}%
    % \includegraphics[width=0.425\linewidth]{figures/sven.png}%
    % \includegraphics[width=0.575\linewidth]{figures/codeql.png}
    % 
    % \includegraphics[width=0.25\linewidth]{figures/codeql_fp.png}
    % \caption{Localization performance of \tool{}'s detection component on CodeQL false positives from our company.}
    \caption{Performance of \tool{}'s detection component on vulnerabilities from SVEN.
    % \ben{SS and I are working on cleaning up this evaluation by increasing the size of the dataset and evaluating single-function metrics.}
    }
    % \vspace{-1em}
    \label{fig:sven-codeql}
\end{figure}

% We report the capabilities of \tool{}'s detection and fix components. Note that we do not emphasize the performance of our detection + fix techniques as the main novelty of our work; we want to show that our implementation of the detection and fix components reaches a baseline of quality and attains performance similar to other state-of-the-art models.
% To make sure that \tool{} can detect and fix components with reasonable false positive and false negative rates, we did probe on the performance of the tool in scenarios close to the real-world use case.
To ensure that \tool{} is both effective and representative of the state-of-the-art, we tested its performance on benchmarks that resemble the real-world deployment scenario as closely as possible.
% 
% \begin{figure}[htbp]
%     \centering
%     \includegraphics[width=0.8\linewidth]{figures/sven_performance.png}
%     \caption{Localization performance of \tool{}'s detection component on the SVEN dataset.\ben{TODO: Remove GPT-3.5 from this figure. We exclude it because it lowers recall a lot and provides lower-quality explanations.}}
%     \label{fig:sven}
% \end{figure}
% 
% 
% Before recruiting developers to try the tool, we thoroughly evaluated \tool{}'s detection and fix capabilities, taking care to make the evaluation settings as close as possible to the tool's use case.
To evaluate \tool{}'s detection capability, we used the SVEN dataset~\cite{sven}, which contains 380 high-quality vulnerability examples from open-source Python projects (93\% label accuracy~\cite{primevul})).
% , and (2) a dataset of 11 fixed and 53 false-positive CodeQL~\cite{codeql} issues reported in X C\# and Java repositories in a major software company.
Our detection model supports all the security vulnerability types present in the dataset.
We used a strict definition for true positives: the predicted bug type must match, and localized line number must match the lines changed in the patch.
\Cref{fig:sven-codeql}  shows on the SVEN dataset our model achieved 80\% Precision and 32\% Recall, with an F1 score of 46\%.
% \wei{You might want to include some state of the art comparsion, say something like, We did not found a vulnerability detection model that handle Web security for Python, PrimeVul paper reported XX F1 fore C vulnerability in SVEN dataset.}
% 
% 
% On the CodeQL dataset, \Cref{fig:sven-codeql} (middle) indicates lower Precision than SVEN \ben{How to explain this? Let's get the entire dataset, then investigate more deeply}.
% , possibly due to the dataset's greater diversity and complexity.
% - possibly because of its diverse distribution and complexity compared to SVEN.
% \Cref{fig:sven-codeql} (right) shows that, compared to CodeQL, our model correctly identified 65\% of true negatives, reducing false positives significantly.

Overall, our results are better than or on par with the prediction quality of SOTA models on vulnerability detection. For example, most recently, \citet{primevul} reported that SOTA models, including CodeBERT, attained 18-21 F1 score on their dataset of C/C++ vulnerabilities. We cannot directly compare our model with other SOTA models on our dataset as most are trained on C/C++-specific memory or pointer bugs~\cite{ivdetect,linevul,deepdfa,reveal,empirical,primevul}.
% ~\cite{codexglue,bigvul,empirical-study,diversevul,primevul}
%due to the large cost of fine-tuning these baseline models on our dataset of vulnerabilities, we cannot compare with these tools on the same dataset. 

\citet{christakis_whatdeveloperswantandneed_2016} found that most developers tolerate up to a 20\% false-positive rate; with the LLM filter, our model meets this threshold on the SVEN dataset, with 80\% precision.
These results highlight \tool{}'s practical effectiveness and potential for deployment in real-world applications.
% Given the strict true-positive criteria and added support features such as explanations and chat, \wei{we believe that staying \tool{} will generate results for AI vulnerability detection tool.}

%these results indicate that our model could be useful to developers.
% \ben{Consider evaluating detection on Vul4J buggy examples to corroborate the SVEN data.}
% Most state-of-the-art models are trained on C++ or a single language, and on different vulnerability types; we . We achieve the same modality of outputs as AIBugHunter~\cite{fu_aibughunter_2023}.
% with the addition of the LLM filter, our model reaches this quality bar, and considering our strict true-positive requirement and the extra support our tool provides to utilize inaccurate predictions, such as explanations and chat interaction, we believe this provides indications that the model could be useful to developers.
% \ben{Baselines are not trained on security vulnerabilities. We achieve the same modality of outputs as AIBugHunter.}
%We achieved feature parity with AIBugHunter~\cite{fu_aibughunter_2023}, while adding explanation and chat capabilities.

To evaluate \tool{}'s fix component, we used the Vul4J~\cite{vul4j2022} dataset, which includes executable tests to reproduce security vulnerabilities.
We assessed the test results, supplemented by manual validation, to verify that the suggested fixes mitigated issues without breaking other functionality.
Among the 24 single-hunk bugs with vulnerability types that our tool handles, our model produced 3 (13\%) correct fixes and 2 (8\%) partial fixes, which resolved the issue but broke 1-3 other tests; 10 (42\%) fixes had errors inserting the generated code into the file and 9 (37\%) fixes could not compile.
% \ben{I will try to investigate some simple fixes for these errors.}.
For efficiency needed for using in IDE, we chose to not run LLM multiple times. These results show that our model performs similarly to SOTA evolution-based automated program repair (APR) tools~\cite{bui_apr4vul_2023} (13\% correct fixes, taking up to 7 minutes in the 75th percentile, intersection $n=24$) and LLMs such as Codex~\cite{wu_dl_apr_2023} (15.4\% plausible fixes on the first try, intersection $n=13$).

%Speed was a major pain point for static analysis tools~\cite{christakis_whatdeveloperswantandneed_2016}, so we chose to avoid running the LLM multiple times.Under these settings and considering the intersection with the set handled by our tool, 

% Note that our purpose of this performance probe is to ensure that \tool{} is a practical tool which can handle real-world vulnerabilities with the performance comparable to the SOTA techniques, and can be a platform to conduct a meaningful study.  It is not mean to be a controlled evaluation in attempt to make a claim that it outperformed the state of the art.

We conducted the above performance probe to confirm that \tool{} can be used to conduct a meaningful study; that it is practical for handling real-world vulnerabilities and offers performance comparable to state-of-the-art techniques. We did not aim for a comprehensive controlled evaluation to claim that \tool{} outperforms the current state of the art.

\section{User Study design}

We developed three research questions to guide our study.

\smallheader{RQ1} Is \tool{} useful in practice?

\smallheader{RQ2} Which aspects of vulnerability detection + fix tools are most useful?

\smallheader{RQ3} What features do developers want from vulnerability detection + fix tools?

\subsection{Study design}

% \wei{key message: We carried out an exploratory case study with a group of 17 developers recruited from a major software company.} \ben{Need to explain why ``exploratory''.}

% \wei{improve the clarity: recruitment, how we run study, how we analyze the data: first, second, }

We carried out an \textit{exploratory case study}~\cite{advanced-ese} with a group of 17 professional developers at Microsoft. We asked users to run \tool{} on projects they were actively developing or were familiar with, and answer survey questions about their perception of the tool. To the best of our knowledge, our tool is the first vulnerability detection + fix tool to be studied in a real-world setting with professional developers on projects which they own. This study enabled us to explore many open questions such as the role of explanations, the developers' tolerance for false-positives or delayed results, and what constitutes an effective fix in a secure development context.
We chose an exploratory study over, e.g. a controlled study, because it elicits rich feedback from developers in a real-world setting. Our approach takes advantage of the developers' deep understanding of their own projects, leading to a more accurate assessment of potential vulnerabilities and providing more valuable insights.
%A controlled study would prevent us from 
% \ben{How did we mitigate bias and add controls to support our conclusions?}
% We designed a two-pronged study to answer our research questions, involving a survey and interview component. These two components of the study are intended to provide complimentary perspectives on different aspects of our questions.

% \textbf{Tool result survey:} We used our tool to scan repositories for security vulnerabilities and asked the developers for their perspectives on the results, including both the detection and fix components.
\smallheader{Recruitment}
We carried out our study with a group of 17 Microsoft developers.
We recruited developers primarily using snowball sampling, with a 53\% participation rate.
% \ben{Move to threats} %Furthermore, we recruited the developers who have diverse levels of experience and backgrounds.
In total, participants scanned a total of 24 projects, 6.9k files, and over 1.7 million lines of source code, and generated 170 alerts and 50 fix suggestions.

\begin{figure}[htbp]
    \centering
\includegraphics[width=\linewidth]{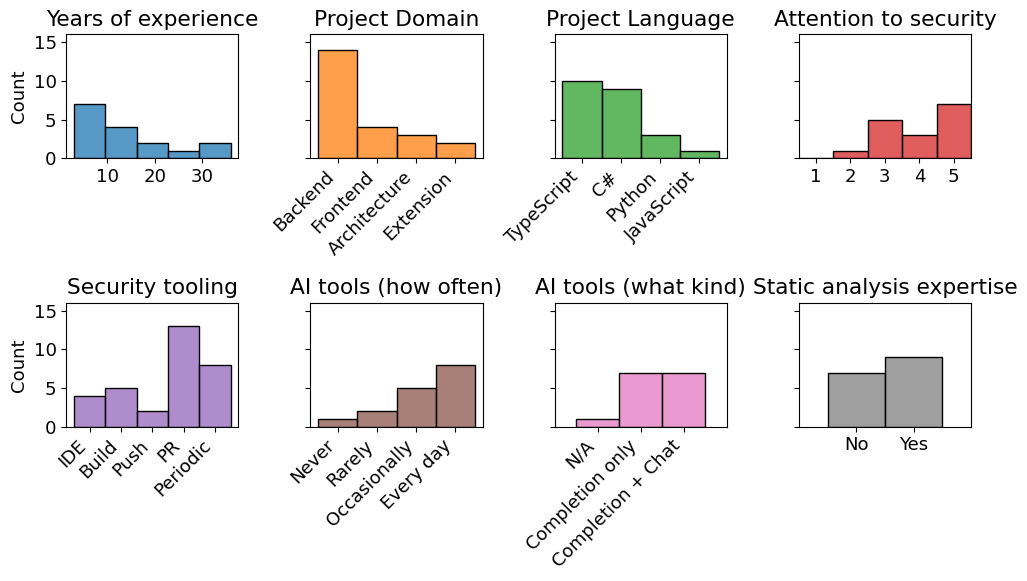}
    \caption{Participant demographics and tool adoption.
    Where applicable, participants listed multiple items for the project domain, project language, and security tooling.
    % \ben{Make sure we factor this information into our data analysis.}
    % \ben{TODO: Add the other 10 participants' data. If it blows up the table, report summary statistics rather than individual data points. Or show a sample of the participants.}}\wei{this can be added in appendix or data package
    }
    \label{fig:demographics}
% \begin{subtable}[t]{\linewidth}
%     \centering
%     \caption{Participant and project profiles.}
%     \label{fig:demographic-user}
%     \begin{tabular}{lllll}
% \toprule
% &&\multicolumn{3}{c}{Project} \\
% \cmidrule{3-5}
% Participant & Experience & Language & Domain  & Application \\
% \midrule
% P1          & 10-20 yrs.      & C\#      & Backend & Data analysis \\
% P2          & 5-10 yrs.      & TS      & Native/Backend & IDE extension \\
% P3          & 5-10 yrs.      & TS      & Backend & Web application \\
% P4          & 20+ yrs.      & C\#, TS      & Frontend/Backend & Web application \\
% P5          & 20+ yrs.      & C\#      & Backend & Data analysis \\
% P6          & 20+ yrs.      & C\#      & Backend & Chat bot \\
% \bottomrule
%     \end{tabular}
% \end{subtable}
%     \quad
% \begin{subtable}{\linewidth}
%     \caption{Participant tool usage.}
%     \label{fig:demographic-tool}
%     \begin{tabular}{lp{1.5cm}p{1.5cm}ll}
% \toprule
% & &&\multicolumn{2}{c}{AI coding tools} \\
% \cmidrule{4-5}
% % \cmidrule(r){3} \cmidrule(l){4-5}
% Participant & Security-consciousness & Security tool usage & Frequency & Usage \\
% \midrule
% P1 & 5/5   & Build/PR & Every day & Completions only \\
% P2 & 3/5   & IDE, PR & Every day & Completions only \\
% P3 & 5/5   & PR & Every day & Completions, Chat \\
% P4 & 5/5   & PR, Periodic & Not at all & None \\
% P5 & 5/5   & IDE, Build, Periodic & Every day & Completions, Chat \\
% P6 & 3.5/5 & Build, IDE & Every day & Completions only \\
% \bottomrule
%     \end{tabular}
% \end{subtable}
\end{figure}

% \smallheader{Demographics}
\Cref{fig:demographics} shows the participants' demographic information (with the exception of one participant who declined the demographic survey), indicating that we studied a diverse set of developers from various levels of experience and backgrounds.
Participants had a median of 11 years of experience; participants worked on both front-end and back-end domains and represented a diverse set of applications such as web applications, back-end services, IDE extensions. Most projects were written in C\# or TypeScript.
Most participants considered themselves security-conscious (median 4 out of 5), and all had some form of security tool running in continuous integration or periodically, though more than half of the projects used tools for reasons of organizational compliance rather than individual initiative; most participants did not use security tools in the IDE.
% 56\% of developers had experience developing static analysis tools.
%which can affect their perception of the tool~\cite{}.
The majority of participants used AI-powered tools occasionally (a few times a week) or every day, such as code completion tools or chatbots, though three participants used AI tools rarely or not at all, stating that they did not find them useful.
% The full demographic survey and data are in our data package.
56\% of developers had expertise in developing static analysis tools, indicating that they were exceptionally knowledgeable about security.

% \textbf{Tool usage interview:} 
\smallheader{Interviews}
% \ben{Cite the total number of alerts that were surfaced and files scanned.}
% Each participant was tasked with \textit{security verification}: they ran our tool on a code-base associated with a production application which they actively develop and shared their perspective guided by both structured and free-form questions.
Each participant ran our tool on a code-base associated with a production application which they actively develop and shared their perspective guided by both structured and free-form questions.
We first asked the participants to run our tool on a simple web server containing a known vulnerability to introduce the features of our tool: detection, fix, and chat.
Then, we directed the participants to run the tool on security-critical areas of their application, such as web interfaces, API endpoints, database code, and file processing code.

We ran the study as a \textit{think-aloud empirical study}~\cite{advanced-ese-think-aloud,ibm-think-aloud}, meaning we asked developers to verbalized their thoughts while running the tool and processing the results.
When participants explicitly asked questions, we provided help and answered questions to facilitate a smooth interview process and clarify the participant's statements, e.g., about the meaning of different UI elements, bug type descriptions, or behavior of the tool, but we refrained from explaining the results of the tool or interpreting the meaning of its outputs to avoid biasing the study.
Each interview lasted approximately 50 minutes, consisting of a 10-minute setup and demographic survey, average 28 minutes usage of the tool and 12 minutes post-usage survey and discussion.
We interviewed all subjects over video calls, and with their full consent, recorded field notes and demographic, audio, screen-capture, survey, and tool usage data.

% \smallheader{Survey}
%, questions shown in \Cref{fig:quantitative-results}
After they used the tool, we asked participants about various aspects of the tool: (1) their overall perception of the usefulness of the detection alerts and suggested fixes and their satisfaction with the speed (Q1-Q3, reported on a Likert scale from 1 to 5 from ``not useful/satisfied at all'' (1) to ``very useful/satisfied'' (5)); (2) whether the tool fits their workflow, whether they trust in the tool, whether the reported alert types were relevant, and whether they would keep using the tool (Q4-Q7, reported as Yes/No).
We also asked the participants what features they found especially useful and what features they would like to see in the tool. We asked the questions verbally during the interview, immediately after trying the tool, in order to collect free-form feedback on each question and ensure that the participant could recollect their experiences with the tool.

We designed the initial set of interview questions, guided by our research questions and informed/inspired by findings and open questions from previous studies of static analysis and AI tools~\cite{johnson_whydont_2013,christakis_whatdeveloperswantandneed_2016,fu_aibughunter_2023,wang_investigatingtrust_2023,smith_whyjohnny_2016,bird_takingflight_2023,nam_llmunderstanding_2024}.
Three authors tried the tool and all authors reviewed the survey questions, and we gathered feedback from outside researchers within our organization to improve the design of the planned questions.
% The full survey and details are in our data package.
% (If sample size is large enough, we'll exclude the first interview as a pilot. Otherwise, let's keep this valuable data.) We conducted a pilot interview which was excluded from the study, which was used to adjust the phrasing and selection of the final survey questions.

% \ben{Clarify that this is just the process, results come later.}
\smallheader{Data Analysis Process}
We analyzed the data quantitatively and qualitatively, reporting the results in \Cref{sec:results}.

% \wei{the following paragraph can be improved with more concrete details. Regarding .... question, we used Yes/no, regarding ... question, we classify the users' response into useful, inserted in correctly..}
% To quantify users' perceptions of our tool, we tally the responses to the post-interview survey and report the average and median Likert scores, or the proportion of Yes/No answers. We also classified the participants' responses to each alert into discrete resolutions, and use these to measure the tool's quality.
To quantify users' perceptions of our tool, we tallied the responses to the post-interview survey, shown in \Cref{fig:quantitative-results}. Regarding questions Q1-Q3, we report the mean and distribution of Likert scores, and regarding Q4-Q7, we report the proportion of ``Yes'' responses.
We also categorized each alert or fix that the participants examined during the interviews into ``Useful'' or one of 8 problem categories, based on the participant's explanation. We discuss the results in \Cref{sec:rq1}.

We conducted a \textit{grounded-theory analysis} to analyze the study participants' rich free-form feedback ~\cite{grounded-theory-analysis}, following the literature~\cite{johnson_whydont_2013,christakis_whatdeveloperswantandneed_2016,picse}.
Grounded-theory analysis is a method used to analyze data by identifying recurring concepts, grouping these concepts into salient categories, and developing themes that provide an overall understanding of participants' perceptions of the tool.
These concepts are derived from participants' quotations, reflecting their thoughts while using the tool and their responses to survey questions.
% \new{This analysis helped us to }
% \wei{cleanup the following wording: which include concepts which occurred frequently in the dialogues, salient groupings of the concepts, and themes, which form a cohesive view of the overall perception of the tool.
% All the codes are driven by quotations from the participants' thoughts while using the tool and responses to survey questions.}
All the resulting concepts and groups are referred to as a \textit{codebook}.
% 151 in codebook excel, 10 verdicts on alert/fix responses

We analyzed over 11 hours of usage and survey transcripts and identified a total of 161 codes in 12 distinct groups. Relevant codes are shown in \Cref{fig:alert-responses} and \Cref{fig:codebook}.
%This process was led by the lead and second authors.
To create the initial codebook, the first and second authors independently analyzed two randomly selected interviews and generated lists of recurring concepts. They then met to create a unified list of concepts, create higher-level groups, and develop overall themes. Each author independently analyzed half of the remaining interviews, periodically syncing and jointly analyzing the same interviews to update the codebook and compare notes.
Both raters agreed on all the classifications for alert responses.
This was an iterative process~\cite{grounded-theory-analysis}, where we created the initial codebook after conducting the first 6 interviews and refactored/added groupings periodically as we conducted the remaining 11 interviews.
% \ben{Measure how much we agreed on the shared transcripts.}
We present our qualitative analysis in \Cref{sec:rq2}.

During the interviews, study participants suggested several features they felt would be useful, which provide useful recommendations for tool builders and directions for further research; we identify these as concepts in our grounded-theory analysis and discuss these feature requests in \Cref{sec:rq3}.
The anonymized demographic data, interview and survey script, and codebook are in our data package~\cite{our-data-package}.

% \begin{itemize}
%     \item Benchmark evaluation: SVEN, internal bugfixes
%     \begin{itemize}
%         \item Setup: The detection model and filter model, the benchmarks, evaluating the detection model on benchmarks, metrics used
%         \item Results: Performance metrics on SVEN and Internal datasets, based on guidelines from ``What do developers want and need''
%     \end{itemize}
%     \item Online evaluation: Scanner reports on internal code
%     \begin{itemize}
%         \item Note any interesting false positives, try to make patterns. I think we might find issues due to lack of context or misleading tokens, but curious to know the extent and find any other patterns.
%     \end{itemize}
%     \item Empirical study: Interviews
%     \begin{itemize}
%         \item Collect telemetry data on what code is analyzed, what features are used, and interview responses
%     \end{itemize}
% \end{itemize}

\section{User study results}
\label{sec:results}

\subsection{RQ1: Is \tool{} useful in practice?}
\label{sec:rq1}

% \ben{Break into sections like this}
% \wei{\bf quantitative results}
% \wei{\bf reasoning why it is not useful, how to improve them}

\begin{figure}[!b]
    \centering
    \includegraphics[width=\linewidth]{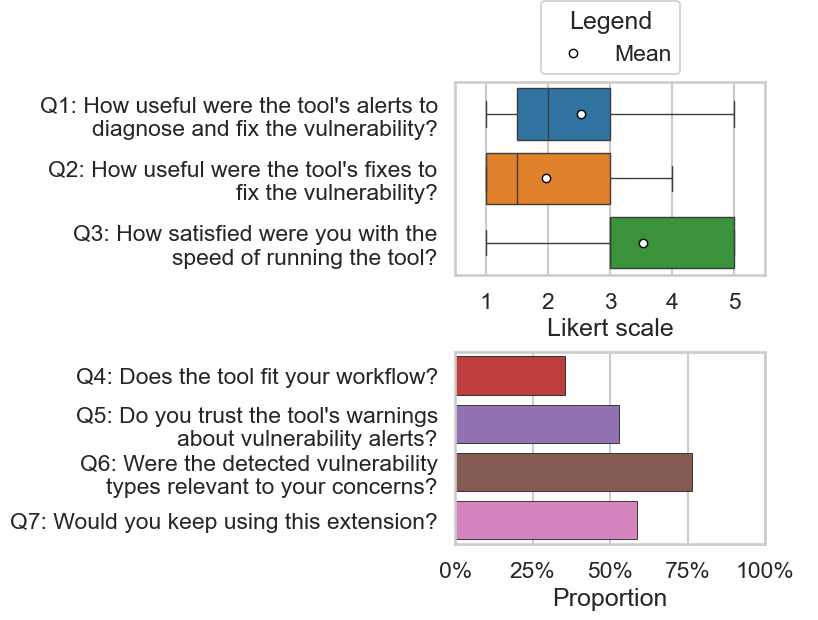}
    \caption{
    Summary of participants' overall perceptions of \tool{}, from our post-interview survey.
    % \ben{Report how many people reported 4/5. Note that the max for detection is 5/5, while max for fix is 4/5.}
    % \ben{Consider adding scatterplot showing individual data points.}
    % \ben{For the no's in Q4-Q7, consider reporting a stacked barplot showing the reasons why not.}
    % \ben{Change the colors of the barplot so they're not shared with the boxplot.}
    % \ben{Consider linking to time, where we developed more features over time which made the tool more useful.}
    }
    \label{fig:quantitative-results}
\end{figure}

% \ben{Usefulness ratings of detections and fixes}
\smallheader{Detection} \Cref{fig:quantitative-results} reports the results of our post-interview survey.
On average, participants rated \tool{}'s alerts at 2.5 out of 5 for usefulness (Q1), with 2 participants giving it a rating of 4.5 or above and 3 participants giving it a rating of 1.
% \wei{Some details can be interesting ---- X people completely satisfied, Y people rate 0/5.}
Only 53\% of participants felt that they trusted the tool's warnings about vulnerability alerts (Q5). \takeaway{The biggest barrier to usefulness and trust in the tool's alerts was the amount of false positives},
with 30\% of users explicitly reporting losing trust in the tool after frequently encountering false positives.
% with \wei{paraphrase it and make it more direct: \ben{XX\%} of users mentioning that after seeing some alerts from the tool, they would be wary of assuming that the next alert shown would be valid.}
% \wei{if we need space, we can take out the false positive comparison with Sven follows}
The false positive rate in real-world settings was higher than in our SVEN dataset measurements (\Cref{sec:benchmarking}).
We attribute this difference to varying languages and vulnerability types: SVEN contains Python code, whereas most participants worked on  Typescript or C\# which comprise only 6\% of our training data.
Additionally, SVEN examples are intra-procedural, lacking information about the calling context and runtime environment, which may widen the gap between benchmark data and real-world testing.

% \ben{Benchmarks are also inside one procedure, }
% \ben{Comment inserted here} %We need to say something about the false positive rate: The false positive rate experienced in the real-world setting was higher than our measurements on the SVEN dataset. We suspect the difference maybe due to the different languages and vulnerability types covered in SVEN vs. the real-world code-based we used for the study. While SVEN data was covering Python, the majority of our participants worked on C\# or Typescript repos. Also, developers usually have a more strict definition of false-positive that may be custom for the code-base they are working on. For example, (do we have an example of a controversial false-positive?)  

%Although participants said the tool showed promise by providing useful alerts and explanations, the usefulness was limited by the number of false positives.  

76\% of participants felt that the vulnerability types detected by the tool were relevant (Q6), with some participants expressing strong approval, for example: \quot{Definitely all of the all the categories of the vulnerabilities that were found here were good. They're all ones that hit these kinds of code all the time.}.

\smallheader{Fix suggestions} On average, participants rated \tool{}'s fix suggestions at 2 out of 5 for usefulness (Q1). 2 participants gave it a rating of 4, and 7 participants gave it a rating of 1.
% \wei{X people rated 5 out of 5, Y people rated 0/5.}
\takeaway{For fixes, one of the most common issues was that the fix was not customized to the developer's codebase}, for example, creating a function to sanitize user inputs when the developer wants to reuse their existing sanitization library; this often prevented the users from directly applying the fix, requiring an overhaul to produce a fix with their intended approach.
This highlights a limitation of common exact match or execution-based metrics for evaluating AI-based fixes, as these metrics do not capture the practical nuances of generating fixes for real-world codebases.

\smallheader{Speed}
The average response time for the tool was 3.9 seconds per file.
More than half of the participants were ``very satisfied'' with the speed of the tool, rating it at 5/5 (Q3). When asked about the tool's speed, one participant stated \textit{``Totally satisfied. I can wait for this kind of stuff''} (referring to security alerts + fixes).

%\ben{Note the average response time for detections + fixes.}
% 
% \ben{Break down some of the common explanations why the usefulness scores were given. Mention blockers for usefulness: inconsistency between fix suggestions and explanations... issues emplacing the fix...}

% \ben{Explanation of why.}

\smallheader{Workflow integration}
More than half (65\%) of participants felt that the tool in its current state would not fit into their workflow (Q4). 14 out of 17 users expressed that {\bf the tool would be more useful if it was running in the background and scanning their code while they were editing or ran along with their build or commit commands}; manually triggering the scan was a barrier to usage, since it required a stopping point in development.
% \ben{Consider removing or specifying that this may drive down/explain the workflow fit stat: Another barrier to workflow was the choice of IDE; XX\% of users did not frequently use VSCode.}

\smallheader{Summary} Although not fully satisfactory, the tool shows promise --- {\bf 59\% of participants expressed that they would keep using the extension.}
% Not enough support
%; additionally, \ben{XX\%} of participants asked to be notified when the full-fledged application is shipped.

%\wei{Trust of the tool}
% \ben{Explanation of why.}
% \ben{Why didn't it fit the workflow? Want automatic scan and think it's redundant with existing tools, this will be rectified by implementing these features and setting the expectation for shifting left to IDE.}
% \ben{What were the reasons why they didn't trust the tool? False positives? Explanations? What are the implications?}
% \ben{The participants who didn't noticed lots of repeated false positive alerts, which weren't super relevant.}
% \ben{What were the specific reasons the users quoted for not wanting to use the tool, or can we connect ``would keep using'' with workflow, trust, and relevance?}

% 
%To study these aspects more deeply, we analyzed the participants' perceptions of each aspect of the tool and features which they thought would improve its usefulness, shown in \Cref{sec:rq2,sec:rq3}.

\begin{figure}[htbp]
    \centering
    \includegraphics[width=\linewidth]{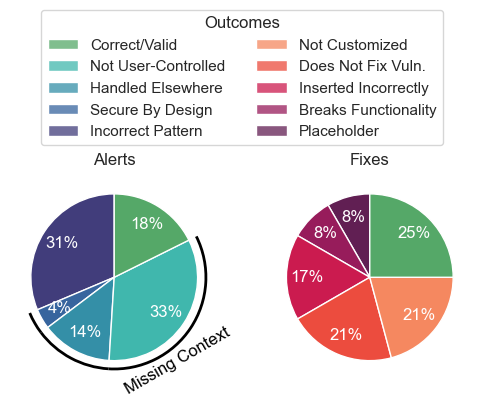}
    \caption{Participant responses to LLM-filtered alerts and LLM-generated fixes while using \tool{}.
    % \ben{Make legend clearer, especially separating alerts and fix.}
    % \ben{Does not fix -- Does Not Fix Vuln.}
    % \ben{Useful -- Correct/Valid}
    % \ben{This is the data from 8 interviews.}
    % \ben{Separate CodeBERT-only alerts from CodeBERT+LLM alerts, or only show the CodeBERT+LLM alerts.}
    % \ben{Merge `does not fix' with `wrong approach' as appropriate and rename `wrong approach' to `improper function choice'.}
    % % \ben{Make this into a pie chart.}
    % \ben{Label FPs with a radius marker.}
    }
    \label{fig:alert-responses}
\end{figure}

% \ben{This section is updated with all the interview data.}
% \ben{TODO: \Cref{fig:alert-responses}, breakdown of quantitative results from participants' responses to the tool's alerts.}
% \ben{This section will be updated to exactly match the figure once the data analysis is complete.}
\Cref{fig:alert-responses} reports the participants' responses to the 51 alerts and 24 fixes for which they provided direct feedback during the interviews,
% \ben{Why did they give feedback on so few? Are there some duplicates in the telemetry used to count the total number of alerts/fixes?},
based on the categories assigned in our grounded-theory analysis. Here we display alerts from the combined CodeBERT + LLM model, excluding four participants who used the CodeBERT model.
% \ben{Need better way to describe this}
% The scan covered 6.9k files and over 1.7 million lines of source code, resulting in 170 alerts and 50 fix suggestions; participants gave direct feedback on 51 alerts and 24 fixes, which are included in the figure. % 49 + 1 revision using copilot chat
Participants considered 18\% of alerts, and 25\% of fixes, to be useful and without significant problems.
% \Cref{fig:good-example} shows an anonymized example of a weak cryptography vulnerability which DeepVulGuard successfully detected and fixed.
\Cref{fig:good-example} shows an anonymized example of a URL redirection vulnerability which \tool{} successfully detected and fixed by adding extra validation. In this case, \tool{} added logic to check that the URL matches a list of approved domains and asked the user to fill the list of domains, based on their knowledge of the application's running context.

\definecolor{pastelyellow}{rgb}{0.99, 0.99, 0.59}

%   public HttpResponse GetRedirectUrl(HttpContext ctx, string url) {
%     var uriBuilder = new Uri(url);
%     var queryParams = ctx.Request.QueryParams;
%     queryParams["redirectCount"] ++;
%     uriBuilder.Query = queryParams;
% -   return Redirect(uriBuilder.ToString());
% +   return SafeRedirect(uriBuilder.ToString());
%   }
% + // Validation stub generated by DeepVulGuard
% + private Response SafeRedirect(string url)
% + {
% +    // TODO: define allowed URLs
% +   List<string> allowedUrls = new List<string> {
% +    "https://www.example.com" };
% +   if (allowedUrls.Contains(url.Split('?')[0]))
% +     return Redirect(url);
% +   else
% +     // TODO: Define default redirect
% +     return Redirect("https://defaultpage.com");
% + }
%   // Get a URL from user input
%   get(string url) {
% +   const allowedUrls = [
% +     "example.com", // TODO: Replace with the actual URL
% +   ];
%     this.notificationService.notify({
%       message: ...,
% -     action: () => { window.open(url); }
% +     action: () => {
% +       if (allowedUrls.includes(url) { window.open(url); }
% +     }
%     })
%   }
\begin{figure}[htbp]
    \centering
\begin{minted}
[
framesep=2mm,
baselinestretch=1.2,
bgcolor=LightGray,
fontsize=\footnotesize,
breaklines,
highlightcolor=pastelyellow,
highlightlines={6,7},
]
{diff}
  function getUrl(string url) {
+   const allowedUrls = [
+     "example.com", // TODO: Provide allowed URLs
+   ];
    this.notificationService.notify({
    // DeepVulGuard: potentially malicious URL redirection.
-   action: () => { window.open(url); }
+   action: () => {
+     if (allowedUrls.includes(url))
+       window.open(url);
+     }
    )
  }
\end{minted}
    \caption{A vulnerability that \tool{} successfully found and fixed by adding validation logic to ensure that an attacker cannot redirect the user to a malicious third-party site.}
    \label{fig:good-example}
\end{figure}

The primary causes of false positive alerts were missing context (totaling 51\% of alerts) and incorrect pattern recognition (31\%). Missing context involved misidentifying variables as user-controlled or overlooking vulnerabilities handled by the calling context or runtime environment.
Incorrect pattern recognition involved misidentifying harmless patterns as vulnerabilities, such as constant strings mistaken for hard-coded credentials.
% \new{The reasons for alert rejections included: the potential vulnerability is already handled by the code, but the detection model lacks context (16.6\%); and false positives, where the code contains no secret, sensitive data, or user-controlled path as the alert claims (XX\%).}
% ; or no vulnerability of the specified type could exist at the alert location (XX\%). We used this feedback, such as the ``obviously incorrect'' cases, to improve the tool.
We hypothesize that incorporating references to the calling context and runtime environment~\cite{rag}, along with in-context examples~\cite{in-context-learning} of commonly misidentified patterns, into the LLM filter prompt shown in \Cref{fig:filter-prompt} may help to address these limitations.
% \ben{Talk about proposals for improving the model.}

\Cref{fig:fp-context-example} shows an example of a ``Not User-Controlled'' outcome. This C\# function's purpose is to redirect the user to a URL retrieved from a database. \tool{} predicted that the field \texttt{resolvedPage.PageURL} could be user-controlled and therefore redirect the user to a malicious third-party website. However, in context, the developer knew that this URL is retrieved from an internally-controlled database, so the URL cannot be overridden by attackers.

% // Redirect the user to a Source in the database, looking it up by its numeric ID.
\begin{figure}[htbp]
    \centering
\begin{minted}
[
framesep=2mm,
baselinestretch=1.2,
bgcolor=LightGray,
fontsize=\footnotesize,
breaklines,
highlightcolor=pastelyellow,
highlightlines={4,5}
]
{csharp}
public HttpResponse RedirectToPage(int pageId) {
  var resolvedPage = Database.LookupById(pageId);
  // 21 lines redacted...
  // DeepVulGuard: potentially malicious URL redirection.
  return Redirect(resolvedPage.PageURL);
}
\end{minted}
    \caption{An example of a false-positive alert caused by missing context. The URL is non-malicious because it is retrieved from an internally-controlled database, not from user input.}
    \label{fig:fp-context-example}
\end{figure}

% -   // Old version directly uses unvalidated input
% +   // Fixed version sanitizes input before usage
\begin{figure}[htbp]
    \centering
\begin{minted}
[
framesep=2mm,
baselinestretch=1.2,
bgcolor=LightGray,
fontsize=\footnotesize,
breaklines,
highlightcolor=pastelyellow,
highlightlines={3,4}
]
{diff}
  app.delete("/room/:roomid", (req, resp) => {
    let roomId = request.params.roomid;
    // DeepVulGuard: potential SQL or script injection.
-   removeRoom(roomId);
+   roomId = sanitizeInput(roomId);
+   removeRoom(roomId);
  });
+ // Sanitization routine generated by DeepVulGuard
+ function sanitizeInput(input) {
+   return input.replace
+     /<script.*?>.*?<\/script>/gi, ""
+   );
+ }
\end{minted}
    \caption{An example of a non-customized fix. In this case, the user would prefer simpler validation, such as checking that \texttt{roomId} is a number, or reusing their project's existing sanitization routines.}
    \label{fig:noncustomized-fix-example}
\end{figure}

\begin{figure*}[t]
    \centering
    \includegraphics[width=\linewidth]{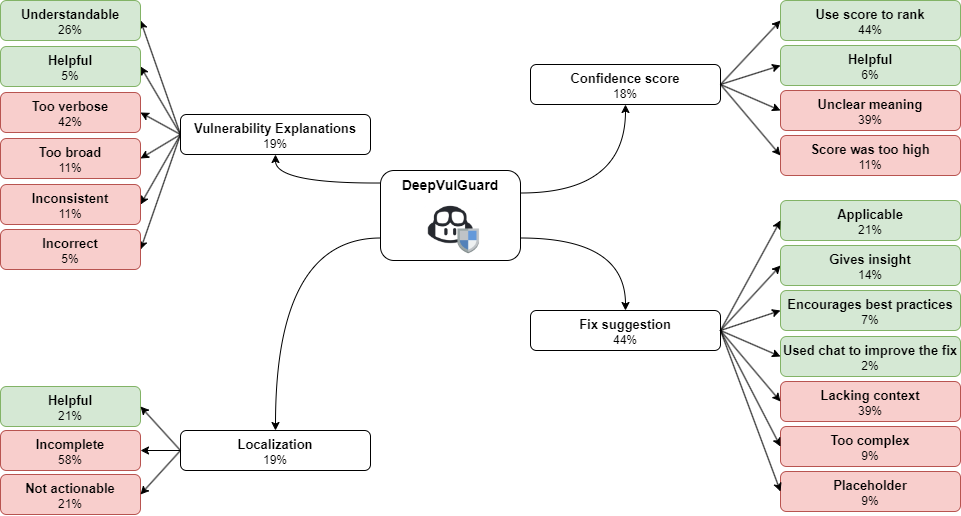}
    \caption{Participants' in-use feedback on the aspects of \tool{}.
    The aspects of the tool are organized into trees, where the leaf nodes are categories of comments from participants and the intermediate nodes are groupings of each category. Percentages show what portion of the comments on their respective aspects that each category constitutes; positive comments are green and negative comments are red.
    % \ben{TODO: Resolve a potential inconsistency between ``directly applicable'' and ``applicable w/ changes'' -- there may be some overlap.}
    % \ben{This is the data from five interviews. The counts are updated but may not be entirely accurate.}
    % \ben{Frequency is counted as the number of occurrences.}
    % \ben{How does this relate to the correctness of the model? Make sure the user knows that the diagram is independent of the accuracy.}
    % \ben{Make fix suggestion one root and syntax one root.}
    % \ben{How many people mentioned each concept?}
    % \ben{Should we compute percentages out of 100\% for each group, or each level?}
    % \ben{Make detection different color, add 100\% there}
    % \ben{}
    % \ben{Add ``Too verbose'' to the Explanations.}
    % \ben{Merge ``Encourages best practices'' and ``Gives insight'' under a new node ``Would apply the fix''.}
    % \ben{Merge ``Not a good starting point'' and ``Incorrect intent'' under a new node ``Would not apply the fix''.
    % Rename ``Valid'' to ``Useful'' and ``Incorrect'' to ``Not useful''.
    % Spread ``Breaks the functionality'' into ``Incorrect intent'' or ``Missing context''.}
    }
    \label{fig:codebook}
\end{figure*}

% Of the fixes which participants didn't want to apply, the most common category was Wrong Approach (25\% of fixes), encompassing cases where the participants thought the fix was too complex, cases where the participant wanted the fix to incorporate the existing functions in the project, and cases where the fix did not comply with the project's style and linting rules.
% 17\% of fixes had incorrect syntax due to incorrectly inserting the code generated by the LLM. 17\% of fixes did not fix the underlying vulnerability, and 8\% of fixes disrupted the existing functionality of the code.
% For 8\% of fixes were placeholders which contained comments suggesting a recommended approach to mitigate the vulnerability but did not directly implement the fix; some participants considered these to be useful, but others would prefer functional fix suggestions which can directly be applied to their code.
21\% of fixes were rejected because they were not customized to the user's codebase -- they did not incorporate existing functions in the project (e.g. sanitization) or didn't comply with project style and linting rules, and thus could not be directly applied.
Additionally, 21\% did not address the underlying vulnerability, another 17\% incorrectly inserted code generated by the LLM, resulting in syntax or indenting issues, and 8\% of fixes mitigated the issue but broke existing functionality. 8\% were placeholders containing instructional comments rather than functional fixes.%; while some participants found these useful, others preferred functional fix suggestions that could be directly applied to their code. \ben{can be deduplicated}
% Finally, one function was considered too complex to apply.
% \ben{Talk about proposals for improving the model.}

\Cref{fig:noncustomized-fix-example} shows an example of a ``Non-Customized'' fix generated by \tool{}. This TypeScript endpoint is intended to look-up a room by its numeric ID and remove it from the database. \tool{} raised a valid concern that the user input could contain malicious values that would allow the user to perform destructive privileged actions, such as a SQL injection. However, the developer would have preferred to simply validate that \texttt{roomId} is numeric, which would catch all possible database or script injections, or reuse one of their project's existing sanitization routines; they stated that this would make the fix easier to understand and maintain.

\subsection{RQ2: Which aspects of vulnerability detection + fix tools are most useful?
% \ben{This section is updated with all the interview data.}
}
\label{sec:rq2}

\Cref{fig:codebook} summarizes the participants' comments about different components of detection and fix tool components, based on the participants' think-aloud feedback while using \tool{}, which we categorized in our grounded-theory analysis.
Based on their responses, fix suggestions and confidence scores seem to be the most useful aspects of the tool, with 44\% and 50\% positive feedback respectively.

\smallheader{Fix suggestions}
% \ben{Fix limitations}
% It's important to have the correct approach for a fix so that the user can implement it with minimal code modifications.
Fixes that have the correct initial approach allow the user to apply them with minimal changes.
Out of the comments on fixes, 21\% noted that the fixes could be applicable, with 7\% of these noting that the fixes required minor changes such as changing variable names or error messages.

\takeaway{Fixes gave developers insight on vulnerabilities and best security practices.}
% \ben{This code should be added to the figure.}
Beyond mitigating the vulnerability, 14\% of comments noted that seeing the diff between `bad' and `good' code helped them understand the root cause of the issue.
Fixes can also provide guidance on secure best practices when developers are working on unfamiliar code (7\%). One developer said, \quot{If I knew I was starting in an area I wasn't very familiar on the most secure practices, it would be very helpful.}

% For security fixes, this involves, among other things, understanding the calling context and environment where the code runs (\wei{In Figure xxx, we reported such mistakes as Missing context}), preserving the intent of the original code (Breaks functionality), and utilizing shared functions in the codebase (Wrong approach).
However, some fixes lacked context (39\%) or were more complex than the user's ideal solution (9\%).
% \new{We reported aspects where our tool was limited in \Cref{fig:alert-responses}, including failing to utilize context (Lacking context, 37\%), disrupting the original intent of the original code (Breaks functionality), and under-utilizing shared functions in the codebase (Wrong approach).}
% \ben{``Wrong approach'' should be renamed to capture intent; all of these are instances of wrong approaches. Maybe rename to ``not customized to codebase''.}
% \ben{Wrong approach is interesting because it's not captured by current benchmarks -- may call for a new metric in evaluation.}
% \ben{parens refer to the pain points called out in the figure.}
% This involves understanding the environment where the code runs \ben{quote ``quote from P6 about dev vs. production environment''}, preserving the intent of the original code \ben{quote from P17 about logging hardcoded secrets}, and utilizing shared functions in the codebase \ben{quote from P4 about utility functions}.
% \ben{Idea for APR metric: measure amount of effort required to modify the suggested code to an acceptable fix.}
Many security issues require multi-site edits, either when changing the semantics of a shared function or when encoding or decoding data; our tool is currently limited to fixing one function at a time, so the fix can lack context, which can result in breaking functionality, expressed by one participant as follows: \quot{This is how [the fix] should be defined, but then I would
have to go find all the places where it's used and fix them all. And that might be a lot of places.}
Incorporating additional context, such as the list of functions which call the function to be changed, could help provide more, contextualized fixes.

% \roshanak{Make the takeaway reflect the new point we're making.}
\takeaway{Placeholder fix suggestions were less preferred for users who were expecting functional fixes.}
LLMs occasionally generate placeholder code by default, including containing instructional comments rather than functional fixes.
% Some users find this useful --
% \quot{I think that's super helpful... If something tells you a problem, it's nice for [the tool] to tell you how to fix it.} --
Some users did not find these useful, constituting 9\% of comments on fixes, such as \quot{Well, that's not helpful} and \quot{It's not really adding anything to the output}.
% \quot{It is putting a helpful comment saying, `make sure we're validating it'... maybe could be more helpful. Maybe instead of just leaving a comment, what to do? Maybe example code or trying to formulate working logic based on what's in the class or the project.}
% \wei{I think this should not a priority, Maybe source an even more negative comment for placeholder fixes? :)}
Since placeholder fixes can negatively impact users who prefer functional fixes, it's important to set expectations; a potential improvement could be to re-generate the fix when placeholders are initially generated, and if a functional fix cannot be provided, the placeholder fix should be accompanied by an explanatory message.

\takeaway{Chat interactions added value by allowing developers to iterate on fixes.}
% \ben{This code should be included in figure 9.}
% \ben{Did only one user use the chat like this?}
In one case where the fix used the wrong approach at first, the developer iterated on the fix by first suggesting a different approach and then specifying their style guidelines, and arrived at a fix which they would apply without having to write the code themselves, saying afterwards, \quot{I like that this is conversational
and I could do a few more rounds of interaction to understand what could be alternative solutions or better ways to approach this issue besides the initial suggestion}.
Before the chat feature was implemented, 50\% of participants expressed the desire to ask the chatbot for more information about alerts or to suggest modifications to fixes.
Recent research supports the potential usefulness of chat interactions. \citet{nam_llmunderstanding_2024} found that developers completed more coding tasks within a given time when using a chatbot for code explanations compared to using a search engine.

% \ben{Open questions, consider adding:}
% \ben{Hypothesis: Users who use chat tools are more likely to pay attention to explanations and chat.}

\smallheader{Confidence score}
\takeaway{Users overwhelmingly used the confidence score to rank issues by importance.}
This interaction constituted 44\% of user feedback on this feature; an additional 6\% noted that the confidence score was helpful for understanding the model's prediction.
% \new{Reporting the score can help the user prioritize alerts by addressing the higher-scored issues first, but may be inappropriate because the confidence score does not indicate the severity of a bug.}
The confidence score can be useful to rank issues, but should be displayed to the user with full understanding of its meaning; 39\% of feedback indicated that participants were unclear about the meaning of the score.
We hypothesize that integrating the severity score~\cite{CVSS} with the model confidence score will make it more useful for prioritizing vulnerabilities.
Finally, a high-confidence result which is a false-positive can degrade the trust in the tool, as seen in 11\% of feedback; therefore, expectations should be managed.
% \ben{\quot{This is wrong to me, although this one gets a higher score.}}.

\smallheader{Vulnerability Explanations}
% \ben{quote from P5 or P8}.
Explaining the vulnerability and security best practices can be useful for providing comprehensive understanding of a vulnerability; 35\% of feedback was positive, indicating that the explanations were understandable and helpful.
One developer compared with existing static analysis tools: \quot{Normally with a static analysis tool, if I get an error that I'm a little unsure on, I would have to go out to a website of track down [an explanation], so providing me a diff and some text here explained it a bit.}
% \ben{One developer said, \quot{If I knew I was starting in an area I wasn't very familiar on the most secure practices, it would be very helpful.}}
% 

\takeaway{With explanations, brevity and adding visual annotations are important}.
42\% of feedback mentioned that the alert descriptions were too verbose; 11\% mentioned that the verbiage was too broad to be useful.
To quote one developer, \quot{If I see the code, it says sanitized input then OK, so I need to sanitize it... I would be more comfortable looking at the fix to know what the issue it is detecting, than read the verbose text.} Later they stated, \quot{I usually read one or two lines and then I stopped there.
If it is too verbose, I probably don't pay too much attention.}. Another developer noted, \quot{Rather than giving me a wall of text, it would be great if it gave me bad and good examples.} One suggestion from this participant is to visually annotate the explanation by presenting labels with short, recognizable names, such as \textit{Path Injection}, and allow the user to read the full explanation if they are interested.
% \ben{If appropriate, insert the quote suggesting this}

\takeaway{Users expect the tool's outputs to be consistent, which introduces challenges when integrating LLM explanations and chat.}
11\% of feedback on explanations noted inconsistencies between the explanation and subsequent fixes.
For example, one user noted that an alert's explanation specified not to use an insecure hashing function \texttt{btoa}, while the suggested fix used this function. While this specific issue could be solved by simply including the LLM-generated explanation in the fix prompt, the general issue is important for tool builders to be aware of.
% \ben{Mention some occurrences of inconsistency between explanation and subsequent fixes/chats (Marked Inconsistent in the figure).}

\smallheader{Localization}
\takeaway{Users preferred highlights on a complete line or variable/string/function call.}
\tool{} highlights the tokens which were localized by the model, which may not necessarily align to semantic boundaries.
21\% of participant feedback noted that localizations was helpful, especially the ability to zoom to a vulnerability' location.
However, 58\% of feedback noted that the localization seemed incomplete because it only highlighted part of the structure it was referencing. This behavior is by-design since the underlying model was specifically trained to only flag parts of the code that contributed to the vulnerability. However, this was one reason that users lost trust in our tool; to quote one user, \quot{The fact that the squiggle starts part way through a word made me wonder -- Oh, is it just on the wrong line, or maybe got some wires crossed somewhere?}.
% 
% \ben{TODO: Rephrase this as a conclusion/takeaway message}
Our model detects vulnerable code patterns at the fault location.
{\bf In 21\% of feedback, users noted that they would prefer to see an alert in a more actionable location -- the root cause, or source for input validation issues, rather than at the fault location}; one user stated, \quot{Preferably, I'd actually do that check way before this, either where we download or where we extract, and that way we know that when we get here, this path is already sanitized.}.
% \new{In general, the region a detection tool highlights in its alerts should align with the user's intent.}

% \ben{Results of RQ2: interview feedback.}
% \ben{Free-form answers, related to the pain points of workflow, false positives, and configuration. \Cref{fig:codebook}.}

% \begin{tcolorbox}
% \smallheader{RQ2 summary} \wei{}
% Confidence scores and explanations were the most useful features for detections.
% Confidence scores were predominantly used to rank issues; high scores on false positives eroded user trust.
% Explanations can aid understanding but should be concise, with short descriptions and diffs improving readability.
% Localizations should align with user expectations, highlighting actionable sections of code and marking complete sections of code.
% % 
% Fixes, even partially correct, provided helpful insights and encouraged best practices. Iterating on fixes was valuable. Placeholder suggestions were divisive, so user configuration options may be beneficial.
% % 
% % Fixes were helpful, even when only partially correct, because they give insight and encourage best practices; some common pitfalls were wrong approach and missing context. It was helpful to allow developers to iterate on fixes. Placeholder suggestions were divisive; some thought they were useful, and others disliked them, so it may be useful to allow users to configure this mode.
% \end{tcolorbox}

\subsection{RQ3: What features do developers want from vulnerability detection + fix tools?
% \ben{This section is updated with all the interview data.}
}
\label{sec:rq3}

\begin{figure}[htbp]
    \centering
    % \vspace{-1em}
    % \includegraphics[width=1.0\linewidth]{figures/feature_requests.png}
    \includegraphics[width=0.9\linewidth]{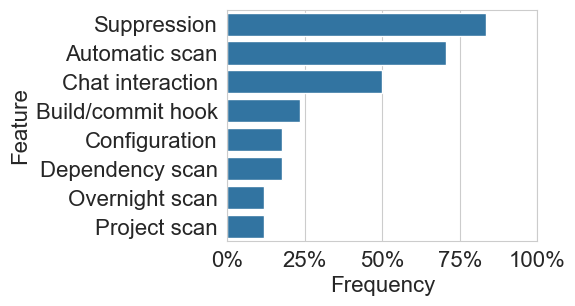}
    \caption{The relative frequency of features suggested by the study participants.
    % \ben{This data comes from 8 interviews. Update after coding all interview data.}
    % \ben{TODO: Clarify the names of the features, such as Explanations or Nightly Run.}
    % \ben{Don't use the light yellow colors.}
    % \ben{Because some participants had an earlier version of the tool, note the percentage of users which requested a feature before it was implemented.}
    % \ben{Count frequency as number of participants instead of number of occurrences.}
    }
    \label{fig:feature-requests}
\end{figure}

% \ben{Quantitative data: Word cloud of the relative frequency of different feature requests.}
\Cref{fig:feature-requests} displays the feature suggestions from the study, identified through our grounded-theory analysis. We implemented basic versions of some highly requested features from early interviews, specifically alert suppression and basic chat interaction. We measured the frequency of these requests from the participants who lacked these features.

\takeaway{Chat interaction was a frequently requested feature.}
Before implementing the chat feature, 3 out of 6 participants wanted a chat interface to better understand alerts, view examples of inputs that triggered them, and refine suggested fixes. Users who had chat available found it useful; for example, one user asked for alternative suggestions to address an alert and expressed, \quot{Yeah, these are good ideas. This would seed some ideas for me}.
% \ben{Add an anecdote about developer using the chat interface, separate from the previous example.}
% Since our tool primarily interacts through a chat window, we added a basic chat feature after six interviews.
% One developer using the chat feature remarked, \quot{I like that this is conversational and I would like to do a few more rounds of interaction to understand what could be alternative solutions or better ways to approach this issue besides the initial suggestion that was provided?}
% \takeaway{Chat interaction was the most frequently requested feature.}
% Participants wanted to interact through a chat interface to better understand the alert, see examples of inputs which would trigger the alert, and make improvements to the suggested fix. While using the early implementation of chat feature in our extension, one developer said, \quot{I like that this is conversational and I would like to do a few more rounds of interaction to understand. You know, what could be alternative solutions or better ways to approach this issue besides the initial suggestion that was provided?}

% Participants also expressed that manually starting the scan would impede their workflow.
\takeaway{Starting the scan manually was a major disruption to developers' workflow.}
12 out of 17 participants expressed that they would prefer if the tool scanned their code in the background and reported problems while they were editing, and 3 participants said it would be useful to trigger scans through a build or commit hook.
One developer explained as follows: \quot{From a workflow perspective, I don't typically reach a point where I say, `Oh, I'm just gonna stop now and go check for security issues'.}
% \ben{We did not implement automatic scanning during the study as it presented a major shift in the usage mode of the tool.}
% % Participants also expressed that manually starting the scan would impede their workflow.
% \takeaway{Starting the scan manually was a major disruption to developers' workflow.}
% One developer expressed that they would like to see the scan happen automatically and surface relevant issues while they're editing, with the explanation: \quot{From a workflow perspective, I don't typically reach a point where I say, `Oh, I'm just gonna stop now and go check for security issues'}.
% XX out of 17 participants similarly expressed that they would prefer if the tool scanned their code in the background and reported problems while they were editing.

% \ben{Take out the second part.}
% \takeaway{Participants wanted the ability to suppress irrelevant alerts, but still be notified if the issue reappears in the future.}
% \takeaway{
% Participants wanted the ability to suppress irrelevant alerts
% , but still be notified if the issue reappears in the future.}
Before we implemented the suppression feature, 5 out of 6 participants wanted the ability to suppress irrelevant alerts. One user specifically mentioned that the tool's outputs would be more manageable if they could filter out alerts identified as false positives or those frequently incorrect or irrelevant to their use case.
% However, they also wanted to be notified if the same issue reappeared due to a code change, as one participant noted: \quot{What I'd like to do is add some kind of comment here to suppress [the issue]... But on the other hand, if something changes around this function, it should check it again and put it in the right context.}
% \takeaway{Participants wanted the ability to suppress irrelevant alerts, but still be notified of recurring issues or changes.}
% Finally, XX out of 17 participants wanted the ability to suppress alerts or suppress all alerts of a certain type and felt the tool's outputs would be more manageable if they could avoid seeing certain types of alerts which were not relevant to their use case or were frequently incorrect.
% They did not want to be bothered by alerts which they had determined they would not fix, although they also wanted to be aware if and when the same issue did reappear -- \quot{What I'd like to do is add some kind of comment here to suppress [the issue]... But on the other hand, if something changes around this function, it should check it again and put it in the right context.}
% 
% 
Participants suggested several other enhancements, such as options for long-running overnight scans, click-to-scan UI interactions, the ability to scan entire directories or projects and their dependencies, and team-shared configurations sensitivity and rulesets.
% \wei{add lightweight explanations for others?}
% \ben{This should be written after all the data analysis is done.}
% \ben{Include in this: Baseline usability features were important, such as click to scan, quick scan/quick fix, and suppressing alerts.}
% However, several participants noted that due to the occurrence of false positives.

% \ben{Recommendation: chat mode with further exploration.}

% \ben{Recommendation: less verbose explanations of alerts. People won't read it.}

% \ben{Recommendation: integrate with the context of the repository to make the fix and detect vulnerabilities.}

% \begin{tcolorbox}
% \ben{TODO: Move these conclusions into ``takeaways'' bolded in the text.}
% \smallheader{RQ3 summary}
% The most common features requested by developers were chat (XX\% of participants), followed by background scanning (XX\%), the ability to suppress alerts (XX\%), and the ability to configuring the tool's ruleset and underlying model (XX\%).
% % 
% Baseline usability features were important, such as click to scan, quick scan/quick fix, and suppressing alerts.
% \end{tcolorbox}

\section{Discussions}
%\section{Lessons Learned}
% Old version in old-lessons-learned.tex
% 
\takeaway{Based on our study, we find that current SOTA AI vulnerability detection and fix tools are not yet satisfactory, but developers remain eager to continue using them.}  This motivates us to continue researching and improving deep learning based methods and tools of vulnerability detection and fix. We have highlighted key lessons from our study in bold in \Cref{sec:results}. Here, we expand on several aspects of evaluating and deploying AI detection and fix tools.

% \smallheader{Recommendations for evaluating AI detection \& fix models}

% Our detection results show a significant difference between benchmark tests and real-world deployment, leading to a much higher false positive rate than expected.
%likely due to a difference in the language and the introduction of inter-procedural vulnerabilities.} 
Our detection results show a substantially higher false positive rate in real-world deployment compared to benchmark tests. Our study indicates it is impractical to (re)train a model for every new code base due to the lack of labeled data. While current test data for AI models often come from the same projects as the training data, in real-world scenarios, AI models are typically applied to unseen projects. Additionally, current deep learning models handle one function at a time~\cite{mbu}, including \tool{}; however, we see that many vulnerabilities in real-world code are related to multiple functions and the runtime environment. Lacking such program and environment context led to 51\% of our tool's alerts being identified as false positives, as shown in \Cref{fig:alert-responses}. This result highlights the need for vulnerability benchmarks that incorporate more realistic contextual information.

We also see that developers usually have specific definitions of ``false positive'' tailored to their own codebase and deployment scenario. For instance, one user dismissed a warning about sensitive data in an SSH key file, citing feasibility issues despite acknowledging the vulnerability:
\quot{This is a problem, but I don't think there's anything they can do about it... I mean, it is a vulnerability -- but if somebody can obtain access to the file system, then they have access to all kinds of password files.}
These user-specific assumptions are difficult to incorporate into dataset labels, emphasizing the need for holistic evaluations in realistic development scenarios.  To improve AI to predict likely feasible bugs, we may need to construct datasets using bugs with reproducible exploits rather than potential (but possibly infeasible) vulnerabilities from CVEs.

%Recent research has shown through benchmark studies that current DL models fail to detect inter-procedural vulnerabilities~\cite{mbu}. Our study confirms and quantifies the impact of this distribution shift in a real-world evaluation scenario.
% Besides the degradation potentially caused by distribution shift of new code, the labels in the benchmark may not agree with what developers think as true/false positives, because they believe they are not likely feasible.Additionally, 

% \ben{This is not cohesive with everything else and is lacking supporting quote; first to cut. \citet{primevul} recently introduced the Vulnerability Detection Score (VD-S) using a 0.5\% false positive rate (FPR) threshold. While this metric is promising, our results suggest developers may tolerate a higher FPR and may lose trust if too few alerts are surfaced. This highlights the need to adopt evidence-based thresholds in evaluation metrics.}

% feasibility to align with the feasibility concerns, we may use data that reproducibable data to train the model
% % 
% \wei{We already learned that vulnerabilities can be related to more than one procedure~\cite{usc-false-positives}. In our study, this is actually very important reason that leads to false positives, making the tool not useful}. 
 
% We hypothesize that evaluation datasets could better align with developer perceptions by utilizing instances of bugs with reproducible exploits rather than CVEs, which indicate potential vulnerabilities but not all of them are exploitable.

We found that 21\% of suggested fixes, though functionally-correct (i.e. they would pass unit tests), were rejected because they were not customized to the user's code-base. Current test execution-based benchmarks~\cite{vul4j2022,defects4j} do not capture this critical issue, highlighting the need for more realistic evaluations that consider this aspect of fix suggestions.
Based on our study, we make several recommendations for deploying AI detection and fix tools. First, when we deployed multiple models for detection, explanations, and fix, we need to ensure that the outputs of these models are consistent, so we do not confuse users. Second, LLM-generated explanations were often too verbose, suggesting a need to guide the LLM to generate concise output and code examples, and to add visual annotations. Third, users prioritize issues based on the displayed score, and this score should reflect important aspects such as severity, not just the confidence score.

\section{Threats to Validity}

Since our work is an empirical study, there may be limits to the generalizability of its findings~\cite{advanced-ese-reporting-results}.

\smallheader{External and internal validity}
% \ben{Limited sample size; mitigate by showing samples from similar studies (johnson et al and fu et al and wang et al), explain there were time constraints and we would have liked more participants, and carried out the analysis until the learnings from new interviews saturated.}
%(over 100k developers)
Our sample of 17 developers from Microsoft may not fully represent all software developers' opinions.
% Literature~\ben{Cite study supporting this, Roshanak will provide} suggests that  8-10 developers are sufficient for feedback on \textcolor{red}{software tools}.
% The same researchers also suggest to break the users in groups and improve the tool based on the findings from each group~\cite{five_users}.
% \ben{Can we leverage this to turn the threat of changing the features into a bonus? We ran the study with a few users, implemented usability features such as directory scan and click to scan, then ran a few and added chat and alert suppression, then the rest.}
Research estimates that 16 users are typically sufficient to fully understand the challenges users face when using a tool~\cite{sample_size}.
This aligned with our findings, as the final two rounds of five interviews each added only 5 and 2 new codes respectively, indicating saturation.
We recruited 11 more developers than a similar user study~\cite{fu_aibughunter_2023}. We included developers with experience ranging from 3 to over 30 years, covering projects in four programming languages and including backend, frontend, and IDE extension code.

Following the literature's recommendation to test iteratively with small user groups~\cite{sample_size_blog}, we first tested with three participants, added key features to our tool like {\it directory scan}, {\it click-to-scan}, and {\it LLM filter}, then tested with three more users before adding {\it chat functionality} and {\it alert suppression}, and finally included the remaining eleven participants.
% \new{To account for this, we present the LLM filter model responses in \Cref{fig:alert-responses} and report feature requests from the developers lacking the requested feature in \Cref{fig:feature-requests}.}
We account for the developing feature set in \Cref{fig:feature-requests} and report only the LLM filter model responses in \Cref{fig:alert-responses}.

We studied one set of models and 27 vulnerability types, which may limit generalization to other models and vulnerability types. The focus of our study was on understanding the practical usefulness of DL models in the IDE, so we chose to study one set of SOTA models (validated in \Cref{sec:benchmarking}).
Our tool supports the top 25 CWEs~\cite{cwe-top-25}, plus the most frequent vulnerability types CodeQL detected in our dataset. Future work could study more models and vulnerability types.

\smallheader{Construct validity}
% In the think-aloud interviews discussed in \Cref{sec:rq2}, developers didn't always give feedback on every tool component~\ben{Can we cite a paper that talks about methodology of running think-aloud studies related to this point?}, and their statements might be biased toward negative feedback aimed at improving the tool~\ben{Can we find any citation that this is likely?}.
We used think-aloud interviews, discussed in \Cref{sec:rq2}, which may result in users providing personal preferences rather than real system issues~\cite{think-aloud-threats}.
We addressed this by using grounded-theory analysis to assess the users' objective verdicts on root causes of alerts and fixes (\Cref{fig:alert-responses}) and quantify the support for each feedback category (\Cref{fig:codebook}).

\section{Related Work}

\smallheader{Deep learning for vulnerability detection and fixing}
Recent research has explored various DL methods for vulnerability detection, including graph neural networks (GNNs)~\cite{reveal,ivdetect,deepdfa} and transformer models~\cite{llmao,linevul,edittime,fu_aibughunter_2023}.
GNNs typically require complete source code to generate the necessary abstract syntax trees (ASTs) and control flow graphs (CFGs), which limits their effectiveness on incomplete code snippets. Our model is based on the state-of-the-art approach from \citet{edittime}, which optimizes for both in-IDE latency and incomplete code snippets.
Compared to \citet{fu_aibughunter_2023}, which uses three separate models for localization, type, and severity prediction, we fine-tune our model to predict the presence, location, and type of vulnerabilities in a single forward pass, enhancing both simplicity and efficiency.
Additionally, we introduce a novel LLM-based filtering technique that improved our model's precision by 20\%; this is compatible with \citet{fu_aibughunter_2023}'s approach.
We also integrate SOTA LLMs for fix suggestions and show that they perform on par with existing DL and APR tools in \Cref{sec:benchmarking}.
% \wei{There are some studies on automically fix software
% you will need to add it to the related work}

% Recent research has proposed various deep learning-based vulnerability detection methods, including graph neural networks (GNN)~\cite{reveal,ivdetect,deepdfa} and transformer models~\cite{llmao,linevul,edittime,fu_aibughunter_2023}.
% Most GNNs require complete source code in order to create the necessary AST and CFG, which loses the potential benefit of scanning incomplete code snippets.
% We built our model following the SOTA approach outlined in \citet{edittime}, since they have optimized for in-IDE latency time and incomplete code snippets.
% Closest to our work, \cite{fu_aibughunter_2023} introduces separate models, utilizing attention weights for localization and separate models for predicting type and severity; we directly fine-tune our model for localization and predict the presence of a vulnerability, localization, and bug type in a single forward pass in order to maximize the simplicity and efficiency of our system.
% Additionally, we introduce a novel LLM-based filtering technique which improved our model's precision by 20\%.
% We incorporate SOTA LLMs for fix suggestions and compared with existing DL and APR fixing tools in \Cref{sec:benchmarking}.

% \ben{Deep learning models for vulnerability detection}
% \ben{Transformers e.g. LineVul, GNN}
% 

\smallheader{Benchmark studies of AI detection + fix models}
% Several empirical studies highlight the importance of evaluating DL vulnerability detection models in real-world scenarios.
Several empirical studies of DL models corroborate our results in underscoring the need for user studies in realistic scenarios.
\citet{reveal} found that DL models often face issues with data duplication and unrealistic distributions of vulnerable classes.
\citet{diversevul} showed that existing models have difficulty generalizing to unseen projects, but increasing the volume of training data can improve their generalization.
\citet{empirical} demonstrated that while some models perform well on benchmarks matching their training data, they may struggle to generalize to new projects and bug types.
Recently, \citet{primevul} indicated that current benchmarks may overestimate the performance of deep learning models.
% These studies corroborate our results in underscoring the need for user studies in realistic scenarios.

% Several empirical studies highlight the need for evaluating deep learning models in real-world vulnerability prediction scenarios.
% The foundational work of \citet{reveal} showed that DL models can suffer from challenges with data duplication and unrealistic distribution of vulnerable classes.
% % We incorporate these findings into our setup for evaluating our models.
% \citet{empirical} showed that, though some models perform well on benchmarks in the same distribution as their training data, they can fail to generalize to novel projects and bug types.
% \citet{diversevul} demonstrated that existing models struggled to generalize to unseen projects and that additional volume of training data can improve the generalization ability.
% Recently, \citet{primevul} showed that existing benchmarks can overestimate the performance of deep learning-based models.
% % \ben{Empirical studies of deep learning-based vulnerability detection models: models perform acceptably well, but can fail to generalize in some contexts, or can be less usable than users would like, e.g. no localization. What are some other findings?}
% % \ben{Empirical study icse2023, diversevul, primevul, \cite{fu_aibughunter_2023}}
% These studies motivate the need for user studies in realistic scenarios. We incorporate these into our findings into our evaluation setup, described in \Cref{sec:benchmarking}.

\smallheader{User studies of static analysis and AI tools}
We developed our tool based on findings and recommendations from several user studies of traditional static analysis tools~\cite{johnson_whydont_2013,christakis_whatdeveloperswantandneed_2016,smith_whyjohnny_2016} (see \Cref{sec:ui} for more details).
A controlled study by \citet{fu_aibughunter_2023} involving six software practitioners demonstrated that DL detection \& fix tools can be beneficial. They also surveyed 21 practitioners about the usefulness of features like localization, type and severity prediction, and fix suggestions, which informed our tool's design.
To our knowledge, we are the first to conduct a study with professional developers on projects they own in a real-world deployment setting. % involving 17 developers, and to explore LLM-generated explanations and chat interactions for vulnerability detection.
% 
% There has been work on investigating whether APR tools are useful in practice, e.g., via user surveys~\cite{survey-apr-1,survey-apr-2,survey-apr-3}.
There has been work on investigating whether APR tools are useful in practice.
Surveys showed that most software practitioners prefer manual bug fixes over current APR tools due to unreliable and slow patch production~\cite{survey-apr-1,survey-apr-2,survey-apr-3}. \citet{apr-live-1} deployed APR in the IDE in a controlled study with 16 developers on a given project; APR increased developers' speed but may impact maintainability. We studied deep learning tools in a real-world development scenario where professional developers run the tool on their own projects. The tool is fast and can improve fixes via conversations with developers.

\section{Conclusions}
Recent research has introduced various deep learning vulnerability tools with promising benchmark performance.
However, there has been no extensive user study on their real-world utility.
To address this, we conducted a comprehensive user study with 17 professional developers, analyzing 24 projects, 6.9k files, and over 1.7 million lines of code, generating 170 alerts and 50 fix suggestions.
% \ben{Emphasize our contributions}
Our study revealed that while current models show promise, they are not yet practical for everyday use due to challenges with (1) false positives caused by missing code context and incorrect pattern recognition, and (2) fixes which were not customized to the codebase.
Based on user feedback, we make several recommendations for aligning model evaluations with real-world development scenarios, and for deploying models in practice.

%emerged as a critical feature for integrating into developers' workflows.

Through our user study, we identified several areas for further research. One direction is to support automatic code scanning and address questions such as when to scan and how much code context to include. Another direction is to further develop AI powered chat interaction. Our preliminary chatbot implementation showed useful for explaining vulnerabilities and generating fixes. Future research should also resolve consistency issues among AI models' outputs.

% \ben{Emphasize our contributions.}

% Not yet full satsifable, but remain very promsing 

% \ben{Key recommendations from our results}

% \ben{Future work:}
% \ben{Scanning while editing. We did not carry out this study because we wanted to see the tool's usage on scanning real projects but would have to do a controlled study to study scanning during edit.}
% \ben{Further study into LLM-generated explanations and chat exploration of the vulnerabilities.}

% Speed of integration with workflow

% \ben{Placeholder} \textcolor{gray}{\lipsum[1]}

%% file: main.bbl
\begin{thebibliography}{54}
\providecommand{\natexlab}[1]{#1}
\providecommand{\url}[1]{\texttt{#1}}
\expandafter\ifx\csname urlstyle\endcsname\relax
  \providecommand{\doi}[1]{doi: #1}\else
  \providecommand{\doi}{doi: \begingroup \urlstyle{rm}\Url}\fi

\bibitem[cop()]{copilot}
{GitHub} {Copilot}.
\newblock \url{https://github.com/features/copilot}.

\bibitem[our(2024)]{our-data-package}
The data package for our study.
\newblock \url{https://doi.org/10.6084/m9.figshare.26367139}, 2024.

\bibitem[wik(2024)]{wikipediaDataBreaches}
{L}ist of data breaches - {W}ikipedia.
\newblock \url{https://en.wikipedia.org/wiki/List_of_data_breaches}, 2024.

\bibitem[Baziuk(1995)]{book2}
W.~Baziuk.
\newblock {BNR}/{NORTEL}: path to improve product quality, reliability and customer satisfaction.
\newblock In \emph{ISSRE}, 1995.
\newblock \doi{10.1109/ISSRE.1995.497665}.

\bibitem[Bessey et~al.(2010)Bessey, Block, Chelf, Chou, Fulton, Hallem, Henri-Gros, Kamsky, McPeak, and Engler]{staticanalysis-coverity}
Al~Bessey, Ken Block, Ben Chelf, Andy Chou, Bryan Fulton, Seth Hallem, Charles Henri-Gros, Asya Kamsky, Scott McPeak, and Dawson Engler.
\newblock A few billion lines of code later: using static analysis to find bugs in the real world.
\newblock \emph{Communications of the ACM}, 2010.
\newblock \doi{10.1145/1646353.1646374}.

\bibitem[Bird et~al.(2023)Bird, Ford, Zimmermann, Forsgren, Kalliamvakou, Lowdermilk, and Gazit]{bird_takingflight_2023}
Christian Bird, Denae Ford, Thomas Zimmermann, Nicole Forsgren, Eirini Kalliamvakou, Travis Lowdermilk, and Idan Gazit.
\newblock Taking {Flight} with {Copilot}: {Early} insights and opportunities of {AI}-powered pair-programming tools.
\newblock \emph{ACM Queue}, 2023.
\newblock \doi{10.1145/3582083}.

\bibitem[Boehm(2002)]{book1}
Barry~W. Boehm.
\newblock \emph{Software Engineering Economics}.
\newblock Springer Berlin Heidelberg, 2002.
\newblock ISBN 978-3-642-59412-0.

\bibitem[Bui et~al.()Bui, Paramitha, Vu, Massacci, and Scandariato]{bui_apr4vul_2023}
Quang-Cuong Bui, Ranindya Paramitha, Duc-Ly Vu, Fabio Massacci, and Riccardo Scandariato.
\newblock {APR4Vul}: an empirical study of automatic program repair techniques on real-world java vulnerabilities.
\newblock \emph{Empirical Software Engineering}.
\newblock \doi{10.1007/s10664-023-10415-7}.

\bibitem[Bui et~al.(2022)Bui, Scandariato, and Ferreyra]{vul4j2022}
Quang-Cuong Bui, Riccardo Scandariato, and Nicol{\'a}s E.~D{\'\i}az Ferreyra.
\newblock {Vul4J}: A dataset of reproducible java vulnerabilities geared towards the study of program repair techniques.
\newblock In \emph{MSR}, 2022.
\newblock \doi{10.1145/3524842.3528482}.

\bibitem[Campos et~al.(2021)Campos, Restivo, Sereno~Ferreira, and Ramos]{apr-live-1}
Diogo Campos, André Restivo, Hugo Sereno~Ferreira, and Afonso Ramos.
\newblock Automatic program repair as semantic suggestions: An empirical study.
\newblock In \emph{ICST}, 2021.
\newblock \doi{10.1109/ICST49551.2021.00032}.

\bibitem[Chakraborty et~al.(2021)Chakraborty, Krishna, Ding, and Ray]{reveal}
Saikat Chakraborty, Rahul Krishna, Yangruibo Ding, and Baishakhi Ray.
\newblock Deep learning based vulnerability detection: Are we there yet?
\newblock \emph{IEEE Transactions on Software Engineering}, 2021.
\newblock \doi{10.1109/TSE.2021.3087402}.

\bibitem[Chan et~al.(2023)Chan, Kharkar, Moghaddam, Mohylevskyy, Helyar, Kamal, Elkamhawy, and Sundaresan]{edittime}
Aaron Chan, Anant Kharkar, Roshanak~Zilouchian Moghaddam, Yevhen Mohylevskyy, Alec Helyar, Eslam Kamal, Mohamed Elkamhawy, and Neel Sundaresan.
\newblock Transformer-based vulnerability detection in code at edittime: Zero-shot, few-shot, or fine-tuning?, 2023.
\newblock URL \url{https://arxiv.org/abs/2306.01754}.
\newblock arXiv: 2306.01754.

\bibitem[Charmaz(2006)]{grounded-theory-analysis}
Kathy Charmaz.
\newblock \emph{Constructing grounded theory: A practical guide through qualitative analysis}.
\newblock Sage, 2006.
\newblock ISBN 0761973532.

\bibitem[Chen et~al.(2023)Chen, Ding, Alowain, Chen, and Wagner]{diversevul}
Yizheng Chen, Zhoujie Ding, Lamya Alowain, Xinyun Chen, and David Wagner.
\newblock {DiverseVul}: A new vulnerable source code dataset for deep learning based vulnerability detection.
\newblock In \emph{RAID}, 2023.
\newblock \doi{10.1145/3607199.3607242}.

\bibitem[Christakis and Bird(2016{\natexlab{a}})]{christakis_whatdeveloperswantandneed_2016}
Maria Christakis and Christian Bird.
\newblock What developers want and need from program analysis: an empirical study.
\newblock In \emph{ASE}, 2016{\natexlab{a}}.
\newblock \doi{10.1145/2970276.2970347}.

\bibitem[Christakis and Bird(2016{\natexlab{b}})]{staticanalysis-microsoft}
Maria Christakis and Christian Bird.
\newblock What developers want and need from program analysis: an empirical study.
\newblock In \emph{ASE}, 2016{\natexlab{b}}.
\newblock \doi{10.1145/2970276.2970347}.

\bibitem[Ding et~al.(2024)Ding, Fu, Ibrahim, Sitawarin, Chen, Alomair, Wagner, Ray, and Chen]{primevul}
Yangruibo Ding, Yanjun Fu, Omniyyah Ibrahim, Chawin Sitawarin, Xinyun Chen, Basel Alomair, David Wagner, Baishakhi Ray, and Yizheng Chen.
\newblock Vulnerability detection with code language models: How far are we?, 2024.
\newblock URL \url{https://arxiv.org/abs/2403.18624}.
\newblock arXiv: 2403.18624.

\bibitem[Distefano et~al.(2019)Distefano, F\"{a}hndrich, Logozzo, and O'Hearn]{staticanalysis-facebook}
Dino Distefano, Manuel F\"{a}hndrich, Francesco Logozzo, and Peter~W. O'Hearn.
\newblock Scaling static analyses at facebook.
\newblock \emph{Communications of the ACM}, 2019.
\newblock \doi{10.1145/3338112}.

\bibitem[Easterbrook et~al.(2008)Easterbrook, Singer, Storey, and Damian]{advanced-ese}
Steve Easterbrook, Janice Singer, Margaret-Anne Storey, and Daniela Damian.
\newblock \emph{Selecting Empirical Methods for Software Engineering Research}.
\newblock Springer London, 2008.
\newblock ISBN 978-1-84800-044-5.

\bibitem[Feng et~al.(2020)Feng, Guo, Tang, Duan, Feng, Gong, Shou, Qin, Liu, Jiang, and Zhou]{codebert}
Zhangyin Feng, Daya Guo, Duyu Tang, Nan Duan, Xiaocheng Feng, Ming Gong, Linjun Shou, Bing Qin, Ting Liu, Daxin Jiang, and Ming Zhou.
\newblock {C}ode{BERT}: A pre-trained model for programming and natural languages.
\newblock In Trevor Cohn, Yulan He, and Yang Liu, editors, \emph{EMNLP Findings 2020}, 2020.
\newblock \doi{10.18653/v1/2020.findings-emnlp.139}.

\bibitem[Fu and Tantithamthavorn(2022)]{linevul}
Michael Fu and Chakkrit Tantithamthavorn.
\newblock {LineVul}: A transformer-based line-level vulnerability prediction.
\newblock In \emph{MSR}, 2022.
\newblock \doi{10.1145/3524842.3528452}.

\bibitem[Fu et~al.(2023)Fu, Tantithamthavorn, Le, Kume, Nguyen, Phung, and Grundy]{fu_aibughunter_2023}
Michael Fu, Chakkrit Tantithamthavorn, Trung Le, Yuki Kume, Van Nguyen, Dinh Phung, and John Grundy.
\newblock {AIBugHunter}: {A} {Practical} {Tool} for {Predicting}, {Classifying} and {Repairing} {Software} {Vulnerabilities}, 2023.
\newblock URL \url{http://arxiv.org/abs/2305.16615}.
\newblock arXiv:2305.16615.

\bibitem[He and Vechev(2023)]{sven}
Jingxuan He and Martin Vechev.
\newblock Large language models for code: Security hardening and adversarial testing.
\newblock In \emph{ACM CCS}, 2023.
\newblock URL \url{https://arxiv.org/abs/2302.05319}.

\bibitem[Humphrey(1995)]{book3}
Watts~S. Humphrey.
\newblock \emph{A Discipline for Software Engineering}.
\newblock Addison-Wesley Longman Publishing Co., Inc., 1995.
\newblock ISBN 0201546108.

\bibitem[{IBM}(2024)]{ibm_cost}
{IBM}.
\newblock {C}ost of a data breach 2024.
\newblock \url{https://www.ibm.com/reports/data-breach}, 2024.

\bibitem[Jedlitschka et~al.(2008)Jedlitschka, Ciolkowski, and Pfahl]{advanced-ese-reporting-results}
Andreas Jedlitschka, Marcus Ciolkowski, and Dietmar Pfahl.
\newblock \emph{Reporting Experiments in Software Engineering}.
\newblock Springer London, 2008.
\newblock \doi{10.1007/978-1-84800-044-5_8}.

\bibitem[Johnson et~al.(2013)Johnson, Song, Murphy-Hill, and Bowdidge]{johnson_whydont_2013}
Brittany Johnson, Yoonki Song, Emerson Murphy-Hill, and Robert Bowdidge.
\newblock Why don't software developers use static analysis tools to find bugs?
\newblock In \emph{ICSE}, 2013.
\newblock \doi{10.1109/ICSE.2013.6606613}.

\bibitem[Johnson et~al.(2023)Johnson, Bird, Ford, Forsgren, and Zimmermann]{picse}
Brittany Johnson, Christian Bird, Denae Ford, Nicole Forsgren, and Tom Zimmermann.
\newblock Make your tools sparkle with trust: The {PICSE} framework for trust in software tools.
\newblock In \emph{ICSE SEIP}, 2023.
\newblock \doi{10.1109/ICSE-SEIP58684.2023.00043}.

\bibitem[Just et~al.(2014)Just, Jalali, and Ernst]{defects4j}
Ren\'{e} Just, Darioush Jalali, and Michael~D. Ernst.
\newblock {Defects4J}: a database of existing faults to enable controlled testing studies for java programs.
\newblock In \emph{ISSTA}, 2014.
\newblock \doi{10.1145/2610384.2628055}.

\bibitem[Keman et~al.(2023)Keman, Wang, Wei, and Madnick]{hbr}
Huangm Keman, Xiaoqing Wang, William Wei, and Stuart Madnick.
\newblock {T}he {D}evastating {B}usiness {I}mpacts of a {C}yber {B}reach.
\newblock \url{https://hbr.org/2023/05/the-devastating-business-impacts-of-a-cyber-breach}, 2023.

\bibitem[Lewis(1982)]{ibm-think-aloud}
Clayton Lewis.
\newblock \emph{Using the ``thinking-aloud'' method in cognitive interface design}.
\newblock IBM TJ Watson Research Center, 1982.

\bibitem[Lewis et~al.(2021)Lewis, Perez, Piktus, Petroni, Karpukhin, Goyal, Küttler, Lewis, tau Yih, Rocktäschel, Riedel, and Kiela]{rag}
Patrick Lewis, Ethan Perez, Aleksandra Piktus, Fabio Petroni, Vladimir Karpukhin, Naman Goyal, Heinrich Küttler, Mike Lewis, Wen tau Yih, Tim Rocktäschel, Sebastian Riedel, and Douwe Kiela.
\newblock Retrieval-augmented generation for knowledge-intensive nlp tasks, 2021.
\newblock URL \url{https://arxiv.org/abs/2005.11401}.
\newblock arXiv: 2005.11401.

\bibitem[Li et~al.(2021)Li, Wang, and Nguyen]{ivdetect}
Yi~Li, Shaohua Wang, and Tien~N. Nguyen.
\newblock Vulnerability detection with fine-grained interpretations.
\newblock In \emph{FSE}, 2021.
\newblock \doi{10.1145/3468264.3468597}.

\bibitem[Meem et~al.(2024)Meem, Smith, and Johnson]{survey-apr-1}
Fairuz~Nawer Meem, Justin Smith, and Brittany Johnson.
\newblock Exploring experiences with automated program repair in practice.
\newblock In \emph{ICSE}, 2024.
\newblock \doi{10.1145/3597503.3639182}.

\bibitem[Microsoft(2024)]{vscode}
Microsoft.
\newblock {V}isual {S}tudio {C}ode - {C}ode {E}diting. {R}edefined.
\newblock \url{https://code.visualstudio.com/}, 2024.

\bibitem[Nam et~al.(2024)Nam, Macvean, Hellendoorn, Vasilescu, and Myers]{nam_llmunderstanding_2024}
Daye Nam, Andrew Macvean, Vincent Hellendoorn, Bogdan Vasilescu, and Brad Myers.
\newblock Using an {LLM} to help with code understanding.
\newblock In \emph{ICSE}, 2024.
\newblock \doi{10.1145/3597503.3639187}.

\bibitem[Nielsen(2000)]{sample_size_blog}
Jakob Nielsen.
\newblock {W}hy you only need to test with 5 users.
\newblock \url{https://www.nngroup.com/articles/why-you-only-need-to-test-with-5-users/}, 2000.

\bibitem[Nielsen and Landauer(1993)]{sample_size}
Jakob Nielsen and Thomas~K. Landauer.
\newblock A mathematical model of the finding of usability problems.
\newblock In \emph{CHI}, 1993.
\newblock \doi{10.1145/169059.169166}.

\bibitem[Noller et~al.(2022)Noller, Shariffdeen, Gao, and Roychoudhury]{survey-apr-3}
Yannic Noller, Ridwan Shariffdeen, Xiang Gao, and Abhik Roychoudhury.
\newblock Trust enhancement issues in program repair.
\newblock In \emph{ICSE}, 2022.
\newblock \doi{10.1145/3510003.3510040}.

\bibitem[{NVD}(2024)]{CVSS}
{NVD}.
\newblock {N}{V}{D} - {V}ulnerability {M}etrics.
\newblock \url{https://nvd.nist.gov/vuln-metrics/cvss}, 2024.

\bibitem[O'Brien and Wilson(2023)]{think-aloud-threats}
Liam O'Brien and Stephanie Wilson.
\newblock Talking about thinking aloud: Perspectives from interactive think-aloud practitioners.
\newblock \emph{Journal of User Experience}, 2023.

\bibitem[OpenAI et~al.(2024)OpenAI, Achiam, Adler, et~al.]{gpt4}
OpenAI, Josh Achiam, Steven Adler, et~al.
\newblock {GPT-4} technical report, 2024.
\newblock URL \url{https://arxiv.org/abs/2303.08774}.
\newblock arXiv: 2303.08774.

\bibitem[Sadowski et~al.(2018)Sadowski, Aftandilian, Eagle, Miller-Cushon, and Jaspan]{staticanalysis-google}
Caitlin Sadowski, Edward Aftandilian, Alex Eagle, Liam Miller-Cushon, and Ciera Jaspan.
\newblock Lessons from building static analysis tools at {Google}.
\newblock \emph{Communications of the ACM}, 2018.
\newblock \doi{10.1145/3188720}.

\bibitem[Seaman(2008)]{advanced-ese-think-aloud}
Carolyn~B. Seaman.
\newblock \emph{Qualitative Methods}.
\newblock Springer London, 2008.
\newblock \doi{10.1007/978-1-84800-044-5_2}.

\bibitem[Sejfia et~al.(2024)Sejfia, Das, Shafiq, and Medvidovi\'{c}]{mbu}
Adriana Sejfia, Satyaki Das, Saad Shafiq, and Nenad Medvidovi\'{c}.
\newblock Toward improved deep learning-based vulnerability detection.
\newblock In \emph{ICSE}, ICSE '24, 2024.
\newblock \doi{10.1145/3597503.3608141}.

\bibitem[Smith et~al.(2020)Smith, Do, and Murphy-Hill]{smith_whyjohnny_2016}
Justin Smith, Lisa Nguyen~Quang Do, and Emerson Murphy-Hill.
\newblock Why can't {Johnny} fix vulnerabilities: A usability evaluation of static analysis tools for security.
\newblock 2020.
\newblock ISBN 978-1-939133-16-8.
\newblock URL \url{https://www.usenix.org/conference/soups2020/presentation/smith}.

\bibitem[Steenhoek et~al.(2023)Steenhoek, Rahman, Jiles, and Le]{empirical}
Benjamin Steenhoek, Md~Mahbubur Rahman, Richard Jiles, and Wei Le.
\newblock An empirical study of deep learning models for vulnerability detection.
\newblock In \emph{ICSE}, 2023.
\newblock \doi{10.1109/ICSE48619.2023.00188}.

\bibitem[Steenhoek et~al.(2024)Steenhoek, Gao, and Le]{deepdfa}
Benjamin Steenhoek, Hongyang Gao, and Wei Le.
\newblock Dataflow analysis-inspired deep learning for efficient vulnerability detection.
\newblock In \emph{ICSE}, 2024.
\newblock \doi{10.1145/3597503.3623345}.

\bibitem[{The MITRE Corporation}(2024)]{cwe-top-25}
{The MITRE Corporation}.
\newblock {CWE} top 25 most dangerous software weaknesses.
\newblock \url{https://cwe.mitre.org/top25/}, 2024.

\bibitem[Wang et~al.(2023)Wang, Cheng, Ford, and Zimmermann]{wang_investigatingtrust_2023}
Ruotong Wang, Ruijia Cheng, Denae Ford, and Thomas Zimmermann.
\newblock Investigating and {Designing} for {Trust} in {AI}-powered {Code} {Generation} {Tools}, 2023.
\newblock URL \url{http://arxiv.org/abs/2305.11248}.
\newblock arXiv:2305.11248.

\bibitem[Winter et~al.(2023)Winter, Bowes, Counsell, Hall, Haraldsson, Nowack, and Woodward]{survey-apr-2}
Emily Winter, David Bowes, Steve Counsell, Tracy Hall, Sæmundur Haraldsson, Vesna Nowack, and John Woodward.
\newblock How do developers really feel about bug fixing? directions for automatic program repair.
\newblock \emph{IEEE Transactions on Software Engineering}, 2023.
\newblock \doi{10.1109/TSE.2022.3194188}.

\bibitem[Wu et~al.(2023)Wu, Jiang, Pham, Lutellier, Davis, Tan, Babkin, and Shah]{wu_dl_apr_2023}
Yi~Wu, Nan Jiang, Hung~Viet Pham, Thibaud Lutellier, Jordan Davis, Lin Tan, Petr Babkin, and Sameena Shah.
\newblock How effective are neural networks for fixing security vulnerabilities.
\newblock In \emph{ISSTA}, 2023.
\newblock \doi{10.1145/3597926.3598135}.

\bibitem[Xie and Min(2022)]{in-context-learning}
Sang~Michael Xie and Sewon Min.
\newblock {H}ow does in-context learning work? {A} framework for understanding the differences from traditional supervised learning.
\newblock \url{https://ai.stanford.edu/blog/understanding-incontext/}, 2022.

\bibitem[Yang et~al.(2024)Yang, Le~Goues, Martins, and Hellendoorn]{llmao}
Aidan~ZH Yang, Claire Le~Goues, Ruben Martins, and Vincent Hellendoorn.
\newblock Large language models for test-free fault localization.
\newblock In \emph{ICSE}, 2024.
\newblock \doi{10.1145/3597503.3623342}.

\end{thebibliography}
